\documentclass[sigconf]{acmart}
\usepackage{graphicx, enumerate}
\usepackage{booktabs}
\usepackage{multirow}
\usepackage{pifont} 
\usepackage{tabularray}
\usepackage{tikz}
\usepackage{xcolor}
\usepackage{float}
\usepackage{multicol}
\usepackage{longtable}
\usepackage{tabularx}

\AtBeginDocument{%
  \providecommand\BibTeX{{%
    \normalfont B\kern-0.5em{\scshape i\kern-0.25em b}\kern-0.8em\TeX}}}

\setcopyright{acmcopyright}
\copyrightyear{2024}
\acmYear{2024}
\setcopyright{rightsretained}
\acmConference[CHI '24]{Proceedings of the CHI Conference on Human Factors in Computing Systems}{May 11--16, 2024}{Honolulu, HI, USA}
\acmBooktitle{Proceedings of the CHI Conference on Human Factors in Computing Systems (CHI '24), May 11--16, 2024, Honolulu, HI, USA}
\acmDOI{10.1145/3613904.3641943}
\acmISBN{979-8-4007-0330-0/24/05}

\begin{document}

\title[The Effects of Group Discussion and Role-playing Training]{The Effects of Group Discussion and Role-playing Training on Self-efficacy, Support-seeking, and Reporting Phishing Emails: Evidence from a Mixed-design Experiment}
\author{Xiaowei Chen}
\orcid{0000-0003-0794-1551}
\affiliation{%
  \institution{University of Luxembourg}
  \streetaddress{11 Prte des Sciences}
  \city{Esch-sur-Alzette}
  \country{Luxembourg}
}
\email{xiaowei.chen@uni.lu}

\author{Margault Sacré}
\affiliation{%
  \institution{University of Luxembourg}
  \city{Esch-sur-Alzette}
  \country{Luxembourg}}
\email{margault.sacre@uni.lu}
  
\author{Gabriele Lenzini}
\affiliation{%
  \institution{University of Luxembourg}
  \city{Esch-sur-Alzette}
  \country{Luxembourg}}
\email{gabriele.lenzini@uni.lu}

\author{Samuel Greiff}
\affiliation{%
  \institution{University of Luxembourg}
  \city{Esch-sur-Alzette}
  \country{Luxembourg}}
\email{samuel.greiff@uni.lu}

\author{Verena Distler}
\authornote{Both authors have contributed equally to this research. Their names are listed in alphabetical order.}
\affiliation{%
  \institution{University of the Bundeswehr Munich}
  \city{Munich}
  \country{Germany}}
\email{verena.distler@unibw.de}

\author{Anastasia Sergeeva}
\authornotemark[1]
\affiliation{%
  \institution{University of Luxembourg}
  \city{Esch-sur-Alzette}
  \country{Luxembourg}}
\email{anastasia.sergeeva@uni.lu}


\renewcommand{\shortauthors}{Chen et al.}
\begin{abstract}
Organizations rely on phishing interventions to enhance employees' vigilance and safe responses to phishing emails that bypass technical solutions. While various resources are available to counteract phishing, studies emphasize the need for interactive and practical training approaches. To investigate the effectiveness of such an approach, we developed and delivered two anti-phishing trainings, group discussion and role-playing, at a European university. We conducted a pre-registered\footnote{Pre-registration link: \url{https://aspredicted.org/qi8ka.pdf}} experiment (N = 105), incorporating repeated measures at three time points, a control group, and three in-situ phishing tests. Both trainings enhanced employees' \textit{anti-phishing self-efficacy} and \textit{support-seeking intention} in within-group analyses. Only the role-playing training significantly improved support-seeking intention when compared to the control group. Participants in both trainings reported more phishing tests and demonstrated heightened vigilance to phishing attacks compared to the control group. We discuss practical implications for evaluating and improving phishing interventions and promoting safe responses to phishing threats within organizations.



\end{abstract}


\keywords{Self-efficacy, Support-seeking, Report phishing emails, Role-playing training, Group discussion, Phishing intervention, Mixed-design experiment, Anti-phishing training}





\begin{teaserfigure}
  \includegraphics[width=\textwidth]{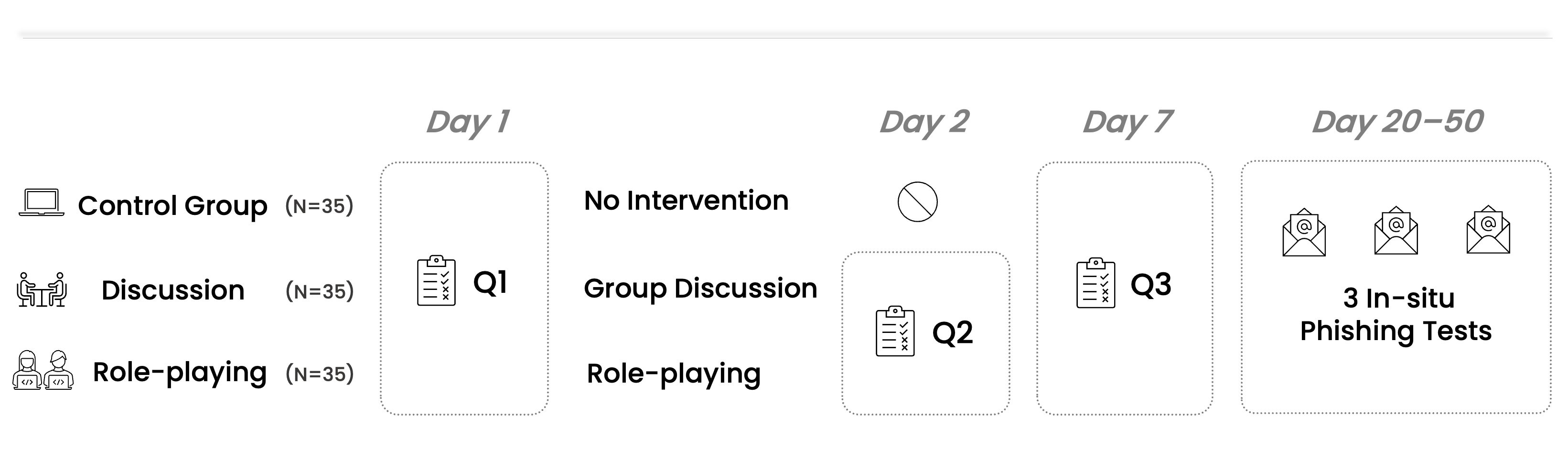}
  \caption{The experimental design: three conditions, repeated measures, and in-situ phishing tests.}
  \Description{Figure 1 illustrates our study design. We had three conditions. A control group with no intervention. Discussion Group employed group discussions as an intervention, while Role-playing Group played the role of hackers as the training. We recruited 35 participants for each condition. We utilized repeated measures and collected questionnaires at three time points. Additionally, we sent three phishing tests to all participants after 20 and 50 days.}
  \label{fig-1}
\end{teaserfigure}

\maketitle

\section{Introduction}
Phishing was the most reported cybercrime between 2019 and 2022 in the US \cite{FBI}. Globally, 4.7 million attacks were recorded in 2022 \cite{APWG}. Phishing attacks exploit human factors by using social engineering techniques to deceive individuals into divulging confidential information or installing malware on their devices \cite{ferreira2015analysis,hatfield2018social}. Phishing is typically used as the initial entry point for more advanced attacks, including ransomware attacks, intellectual property theft, and business scams \cite{goel2018mobile}, which can result in billions of dollars annual losses for organizations \cite{FBI2}. As communication channels like email, instant messenger, and team collaboration tools become popular, attackers are exploiting new vulnerabilities to target technology users \cite{hatfield2018social,aleroud2017phishing,Microsoft}. 


Organizations employ multiple technical measures to reduce the number of phishing attacks, but these measures may fail \cite{Tellinguser}. In such cases, employees' vigilance against phishing can serve as the last line of defense for organizations \cite{mansfield2017raising,zimmermann2019moving}. Employees' awareness and proactive responses to phishing attacks improve organizational information security \cite{hillman2023evaluating,bayl2022response}. Employees who are aware of their organization's information security policies and procedures demonstrate greater competence in managing cybersecurity tasks than those who are not \cite{li2019investigating}. In addition, employees’ self-efficacy in information security positively influences their intentions to comply with security policies \cite{vance2012motivating}. Consequently, organizations implement a range of phishing interventions to heighten employees' awareness and develop their competence in countering phishing \cite{franz2021sok}. 

In addition to on-site/online education, organizations send simulated phishing emails to monitor employees' responses and resistance to phishing attacks (referred to as \textbf{simulating phishing tests}) \cite{jansson2013phishing,dodge2007phishing}. While simulated phishing tests have been commonly adopted by organizations to raise employees' phishing awareness, they alone may not adequately train users to respond safely \cite{volkamer2020analysing}. Accordingly, the \textbf{embedded phishing campaign} was developed, in which employees who respond unsafely to the simulated phishing email are directed to a webpage containing educational resources. Studies on the effectiveness of embedded phishing campaigns in training employees for safe responses have yielded mixed results \cite{kumaraguru2009school,yeoh2022simulated,lain2022phishing,de2020real}. Given the evolving nature of phishing attacks and the critical role of employee vigilance \cite{greitzer2021experimental}, studies suggest developing more practical and interactive anti-phishing training for employees \cite{jampen2020don,williams2018exploring,Tally23}.

Role-playing training helps participants gain experience with various situations, equipping them with the skills and knowledge to anticipate, adapt to, and recover from undesirable situations \cite{jones1987participatory,woltjer2006role}. A recent role-playing anti-phishing training, ``What.Hack'', was found to be effective in improving users' performance of identifying phishing emails within a controlled laboratory setting \cite{wen2019hack}. This effectiveness was evident when compared to the results of two alternative training programs \cite{wen2019hack}, but engaging with What.Hack does not require social interactions between users and the contextual factors embedded within the game narrative might potentially lack relevance in the work context \cite{wen2019hack}. Given that previous research has shown that contextual factors affect employees' susceptibility to phishing attacks \cite{frank2022contextual} and workplace social interactions can help employees protect themselves against such attacks \cite{distler2023influence}, it is imperative that we incorporate these aspects when designing anti-phishing training programs.   

To further explore practical and interactive training approaches tailored for the work context, we have specifically designed two anti-phishing trainings: group discussion and role-playing. Both trainings are designed with the goal of enhancing employees' \textit{anti-phishing self-efficacy} and \textit{support-seeking intention}. They aim to train employees to respond safely to phishing emails. During the group discussions, participants are encouraged to \textbf{share their experiences and anti-phishing practices} with their colleagues \cite{safa2016human}. In the role-playing training, we guide participants to \textbf{think like a hacker} and design phishing emails to infiltrate the organization \cite{esteves2017improve}. By comparing these two training approaches, we intend to address the following research questions (RQ):

\begin{itemize}
    \item RQ1: What is the effect of role-playing training on employees’ anti-phishing self-efficacy compared to group discussion and the control group?
    \item RQ2: What is the effect of role-playing training on employees’ support-seeking intention when receiving phishing emails compared to group discussion and the control group?
    \item RQ3: How do group discussion and role-playing training influence employees' response to phishing attacks? 
\end{itemize}

To address these questions, we employed a \textbf{mixed-design experiment} \cite{charness2012experimental}, assessing training effects both within and between subjects. We incorporated repeated measures at three time points, included a control group, and conducted three simulated phishing tests in our study. This method is novel in that it allows us to assess the effectiveness of anti-phishing trainings with respect to participants' self-efficacy, support-seeking intention, and responses to phishing emails. This paper makes four primary contributions:

\begin{itemize}
    \item We contribute to the understanding that group discussion and role-playing are effective anti-phishing training approaches that enhance employees' perceived self-efficacy, support-seeking intention, and vigilance towards phishing attacks.
    \item Our study highlights the significance of discussing phishing incidents and anti-phishing practices in the workplace, demonstrating its potential to promote safe responses to phishing attacks.
    \item We introduce a new and useful measurement for evaluating anti-phishing trainings: support-seeking intention when receiving suspicious emails.
    \item Our study is one of the first to employ a mixed-design experiment to assess the effects of anti-phishing trainings in the field, measuring both self-reported and behavioral changes. We demonstrate that it is possible to obtain informed consent from research participants for simulated phishing tests and still receive useful insights from the results. 
\end{itemize}

Reflecting upon the results of our study, we advocate for role-playing as an enjoyable and effective approach to engage employees with anti-phishing training. Our study is among the first to demonstrate the effectiveness of two trainings aimed at increasing employees' propensity to report phishing emails through in-situ phishing tests over a time period of three weeks. 

\section{Related Work}
\label{sec-relatedwork}

In the following subsection \ref{phishing-edu}, we review previous studies on anti-phishing education and training\footnote{Anti-phishing education and training: Following the phishing intervention taxonomy \cite{franz2021sok}, we examine studies related to \textbf{phishing education} (which involves developing knowledge and understanding of phishing) and \textbf{training} (aimed at cultivating skills that users can apply when encountering phishing). Refer to the Supplementary Material for an overview of the role-playing and other types of anti-phishing trainings reviewed in our study.}. Then, we review the methods that have been employed to evaluate educational and training interventions in subsection \ref{phishing-method}. Lastly, we examine the role of self-efficacy and social interaction in countering phishing attacks in subsection \ref{phishing-efficacy}.  

\subsection{Anti-phishing education and training}
\label{phishing-edu}


Simulated phishing campaigns and on-site/online education are among the popular interventions adopted by organizations to bolster their phishing resilience \cite{jampen2020don,yeoh2022simulated,kumaraguru2008lessons,gopavaram2021cross}. For organizations with a large number of employees, simulated phishing campaigns seem to be a convenient solution to raise phishing awareness \cite{dahabiyeh2021factors}. Longitudinal observations from an Australian educational institute revealed that employees exhibited safer responses after six cycles of embedded phishing campaigns, as opposed to the period when only simulated phishing tests were administered \cite{yeoh2022simulated}. However, another large-scale and long-term study in Switzerland found contradicting results and argued that embedded training during simulated phishing tests does not make employees more resilient to phishing \cite{lain2022phishing}. Besides these mixed results, deploying phishing campaigns is expensive and requires dedicated human resources in organizations \cite{brunken2023properly,rizzoni2022phishing}. Instead of off-the-shelf phishing campaigns, some organizations have designed their own training programs for employees; for example, 409 employees of a German organization improved their skills in distinguishing phishing email screenshots significantly after attending on-site tutorials \cite{reinheimer2020investigation}. When comparing instructor-, computer-, and text-based anti-phishing training based on the same content, Stockhardt et al. found that instructor-based training was more effective in transferring knowledge than the other two formats and had the highest scores in user satisfaction and confidence \cite{stockhardt2016teaching}. In an experiment conducted within an organization, participants who underwent an adversarial training, adopting the mindset of a cybercriminal, revealed a nearly threefold decrease in susceptibility to phishing attacks compared to those who received a video training \cite{becker2023phishing}.

Role-playing has been a central approach employed by several anti-phishing digital and card games to engage users in learning \cite{zhang2021systematic}. Digital games, including Anti-phishing Phil \cite{sheng2007anti}, Bird’s Life \cite{weanquoi2018using}, and What.Hack \cite{wen2019hack}, have been created to teach users how to identify phishing elements. These games have primarily been evaluated with university students \cite{wen2019hack,weanquoi2018using}; therefore, the effectiveness of these trainings for organizational employees require further empirical investigation \cite{hart2020riskio,beckers2016serious}. On the other hand, card games, which often incorporate red team (attackers) and blue team (defenders) designs, have also been introduced to assess the vulnerabilities posed by social engineering attacks \cite{beckers2016serious,hart2020riskio} and raise awareness of excessive online information disclosure, enhancing phishing awareness \cite{fatima2019persuasive}. Card games appear to be more accessible than digital games, but they tend to exhibit a level of complexity that can challenge users' ability to quickly grasp the game rules \cite{baslyman2016smells}, necessitating additional learning efforts and expert guidance \cite{fatima2019persuasive}. 

\subsection{Evaluating educational and training interventions}
\label{phishing-method}

The majority of our reviewed studies conducted either user evaluation \cite{hart2020riskio,weanquoi2018using,fatima2019persuasive}, which focuses on studying the usability of the intervention, or between-subjects experiment \cite{wen2019hack,hull2023tell,becker2023phishing}, which compares the training effects with alternative interventions; only one study chose the approach of measuring the training effects over time in the field \cite{reinheimer2020investigation}. Post-training questionnaires were frequently used to assess participants' learning experience \cite{wen2019hack}, self-efficacy \cite{hull2023tell}, perceived effectiveness \cite{hart2020riskio}, and feedback \cite{sheng2007anti}. A few studies compared pre- and post-training questionnaires to evaluate training efficacy. The measurements include phishing knowledge \cite{weanquoi2018using}, confidence level \cite{cj2018phishy}, behavioral tendencies, self-reported computer skills and perceived risks \cite{sumner2022examining}. Further, participants' demographic information was commonly collected to examine their relationship with phishing intervention outcomes \cite{fatima2019persuasive,carella2017impact,lain2022phishing}.

Log data of simulated phishing tests and participants' performance in distinguishing phishing emails/websites from legitimate ones have been used as indicators of training efficacy in many previous intervention studies \cite{chaudhary2022developing}. The reporting rate and click-through rate of phishing campaigns have been used to evaluate participants' responses to phishing emails \cite{dodge2007phishing,yeoh2022simulated} and to evaluate the effectiveness of different education approaches \cite{carella2017impact,lain2022phishing,Washtraining}. Through online surveys, Reinheimer et al. measured employees' performance of distinguishing phishing email screenshots from legitimate ones \cite{reinheimer2020investigation}. In the laboratory, participants were instructed to classify emails/websites as phishing or legitimate ones before and after the intervention to assess their effectiveness \cite{kumaraguru2010teaching,sheng2007anti,wen2019hack}. In a literature review \cite{chaudhary2022developing}, Chaudhary et al. compared existing evaluation methods and considered simulated phishing tests provide more realistic view of participant's responses than question-based tests, assuming they comply with data protection laws and are conducted in ethical ways.


Furthermore, observation of time spent on the task \cite{baslyman2016smells}, group discussions \cite{beckers2016serious}, and participants' designs \cite{fatima2019persuasive} have been analyzed to evaluate the interventions. In an in-situ deception study, Distler combined observation and interview data to study employees' responses in the context of their typical work tasks to spear phishing attacks, including social interactions, and their reporting behavior as well as rationalizations \cite{distler2023influence}. In-situ studies evaluating educational and training interventions deliver important insights as they maximize ecological validity \cite{reinheimer2020investigation,chaudhary2022developing}, enable researchers to observe how context influences reactions to a social engineering attack, and allow for the capture of natural reactions at the critical moment when an employee is exposed to a social engineering attack. However, many ethical challenges are associated with conducting ecologically valid phishing studies; for a discussion refer to \cite{resnik2018ethics}. 


\subsection{Self-efficacy and social interaction in anti-phishing}
\label{phishing-efficacy}

Self-efficacy is the most frequently studied construct from Protection Motivation Theory when applied to users' information security behaviors \cite{haag2021protection}. Self-efficacy in information security is defined as \textit{a belief in one’s capability to protect information and information systems from unauthorized disclosure, modification, loss, destruction, and lack of availability} \cite{rhee2009self}. Studies show self-efficacy positively influences employee's information security compliance intention \cite{menard2017user,herath2009protection,vance2012motivating}. Specifically in anti-phishing studies, self-efficacy positively influenced an individual's likelihood of reporting phishing emails \cite{kwak2020users,marin2023influence} and mobile users' motivation to avoid phishing \cite{verkijika2019if}. Employees with stronger self-efficacy were more inclined to share their negative experiences with colleagues, alerting them about phishing attacks in a financial company \cite{conway2017qualitative}. In this study, we define \textbf{anti-phishing self-efficacy} as a belief in one's ability to recognize suspicious emails and keep up to date with phishing techniques \cite{williams2020developing,ng2009studying}. Several self-efficacy scales have been developed to measure users' self-efficacy in domains such as information security \cite{rhee2009self} and smart home security \cite{borgert2023home}; however, there is currently no comparable scale available for measuring anti-phishing self-efficacy. 

Promoting social interaction within organizations can benefit the organization in defending itself against phishing threats. Firstly, workplace social interactions can support employees in assessing and responding to suspicious emails \cite{distler2023influence}. By motivating employees to report suspicious emails, organizations can detect phishing attempts within minutes after a new attack is launched \cite{lain2022phishing}. Secondly, Das et al. found that ``social cues'', where individuals engaged with or observed others' actions, were the predominant category of triggers that led to recent security and privacy behaviors in an online survey (N = 852) \cite{das2019typology}. Thirdly, stories shared by peers, second only to expert advice \cite{Washtraining}, led to lower click rates in phishing tests compared to those who received other forms of training materials in two experimental studies \cite{marsden2020facts}. Thus, research suggests that promoting social interaction and experience-sharing at the workplace holds promise as an effective approach for defending against phishing attacks.

\section{Methods}
\label{sec-method}

\subsection{Group discussion and role-playing training design}
\label{design}
For the purpose of the study, we developed two training programs: a group discussion and a role-playing training. The underlying mechanism behind using group discussion to train employees stems from a study showing that small group discussions can bring about powerful and lasting changes in information security awareness and behavior \cite{albrechtsen2010improving}. The role-playing training design was inspired by two previous studies showing that role-playing is an engaging approach to train anti-phishing skills \cite{wen2019hack} and that by assuming the role of hackers, students improve their awareness of spear phishing risks \cite{fatima2019persuasive}. We chose a face-to-face training approach, instead of a digital format, to foster more social interaction between participants \cite{gutfleisch2022putting}.

To facilitate the comparison between group discussion and role-playing training, we used the same set of materials to design them: recent real phishing emails targeting the organization, a phishing definition \cite{ferreira2015analysis}, content cues \cite{sergeeva2023we,ferreira2015principles}, attack channels, and phishing techniques \cite{aleroud2017phishing}. In the process of designing both training programs, we consulted two experts from the organization's Information Security Office with expertise in phishing and three professors working on cybersecurity and Human-computer Interaction. Prior to data collection, we conducted a pilot study of the role-playing training with 7 employees to gather feedback and refine the training procedure. The group discussion and role-playing training shared the same structure and length; they both started with a brief introduction of the study and training schedule, a tutorial on phishing fundamentals, group discussion or group work, and a conclusion, as outlined in Table \ref{table-1}. 

\begin{table*}[!ht]
\centering
\caption{Group discussion and role-playing training procedure.}
\label{table-1}
\begin{tabular}{|c|c|c|c|}
\hline
       & \multicolumn{2}{c|}{Group discussion} & Role-playing            \\ \hline
5 min  & \multicolumn{3}{c|}{Introduction}                                         \\ \hline
20 min & \multicolumn{3}{c|}{Phishing fundamentals: definition, content cues, channels, and techniques}                                \\ \hline
\multirow{3}{*}{50 min} &
  \multicolumn{2}{c|}{\multirow{3}{*}{\begin{tabular}[c]{@{}c@{}}Analyze real phishing emails with a template (10 min). \\
  Discuss phishing emails, perceived vulnerability, \\ 
  and coping strategies (40 min). \end{tabular}}} &
  \multirow{3}{*}{\begin{tabular}[c]{@{}c@{}}Group work on real phishing emails and \\ 
  design one phishing email (40 min). \\
  Phish each other and identify the phishing email (10 min).\end{tabular}} \\
       & \multicolumn{2}{c|}{}                      &                              \\ 
       & \multicolumn{2}{c|}{}                      &                              \\ \hline
10 min & \multicolumn{3}{c|}{Conclusion phase}                                       \\ \hline
\end{tabular}%
\end{table*}

In the group discussion condition, we first asked each participant to scrutinize two real phishing emails with a template with questions on suspicious elements of the emails and the difficulty of identifying them as phishing. Then, we moved to discuss the following questions in groups of 4 to 6 people facilitated by a researcher:

\begin{itemize}
    \item \textit{What surprised you most about these real phishing emails?
    \item Have you received phishing emails on your work accounts?
    \item How do you respond to these suspicious emails related to work?
    \item Which contextual factors related to your job position might be exploited by attackers?
    \item What would happen if you were to click on a suspicious email or download malware to your work laptop?}
\end{itemize}

In the role-playing condition, we asked participants to play the role of hackers aiming to infiltrate the organization. We randomly divided participants into two groups, each comprising 2-4 individuals. Each group was equipped with a computer, an email account of a fictitious persona, and two legitimate work-related emails in their draft box. The group work started with a discussion on suspicious elements and attack techniques of real phishing emails and, subsequently, creating one phishing email together to phish the other group. 40 minutes later, the participants sent the created phishing email and two legitimate emails to the other group. After both groups sent their ``phishing'' email and legitimate emails, they were asked to identify the phishing email created by the other group.

In the conclusion phase, participants from both conditions shared strategies and practices they intend to use to protect their workplace from future phishing attacks, as well as the lessons they learned during the training. Additionally, we provided participants with additional tips for identifying phishing emails \cite{NIST} and phishing awareness resources available at the organization.

\subsection{Participants}

We employed multiple methods to recruit participants for our study. We sent study invitations via email to all employees across four university faculties (Humanities, Engineering, Computer Science, and Medicine), posted recruitment materials on campus in the form of printed posters and digital displays, distributed flyers in two employee cafeterias, and conducted door-to-door recruitment in two office buildings (we include the recruitment poster in the Supplementary Material). 118 employees registered interest to participate in the study by completing an online questionnaire indicating their faculty, email address, and availability. We did not exclude any participants, and any current employee who had a work email account was allowed to participate in our study. We invited all employees who expressed interest to participate in our study.  

Among the 105 employees who participated in our study, 60 reported being female, 40 male, one non-binary, and four chose not to disclose their gender. 60\% (N = 63) of participants were aged between 25 and 34, 21\% (N = 22) were between 35 and 44, and 13\% (N = 14) were between 45 and 54. In terms of their professional background, 47 were from the Humanities faculty, 22 from Engineering, 9 from central administration, 9 from Medicine, 8 from Computer Science,  and 10 from other departments. 39 participants worked as doctoral researchers, 15 as postdoctoral researchers or research scientists, and the remaining participants held roles as research facilitators, Research \& Development specialists, administrators, and professors. Their work experience at the current organization varied between 1 and 187 months (mean = 42.6, SD = 46.6). 

\subsection{Study procedure}

\subsubsection{Conditions}

Participants were randomly assigned to one of three conditions to compare the effects of both trainings to a control condition\footnote{We requested participants to indicate their availability among the 20 provided timeslots in the registration form. Timeslots that reached 8 or more participants were designated for group discussion or role-playing training randomly. For timeslots that had fewer than 8 participants, we assigned them to the control group.}. The \textbf{control group} (N = 35) received no intervention from us and was conducted remotely. The \textbf{group discussion} (N = 35) and \textbf{role-playing} (N = 35) training sessions took place in the same user lab. We invited participants to attend a ``Phishing Resilience Workshop'' in groups, but they were unaware that there were two different training programs. 

\subsubsection{Pre-training assessment}

After we assigned registered participants to three conditions, we sent them study invitations with an attached information sheet and consent form for the study. We included a link to the first questionnaire (\textbf{Q1}) in the invitation. For the control Group participants, we instructed them to reply with a signed consent form to participate in the ``Phishing Resilience Study''. They could answer Q1 immediately after they sent us the consent form. For participants of two treatment conditions, we instructed them to fill out Q1 prior to their attendance of the training session.

\subsubsection{Post-training assessment}

Immediately after each group discussion and role-playing training session, we sent the second questionnaire (\textbf{Q2}) to the session attendees and asked them to complete Q2 within 24 hours. The control group participants did not receive Q2 because they had not taken part in an activity.

On the seventh day after each training session, we sent the attendees the third questionnaire (\textbf{Q3}) to measure knowledge retention and gather feedback on the training. Additionally, on the seventh day after the control group participants answered Q1, we sent them Q3 with adapted questions.

\subsubsection{In-situ phishing tests}

To assess the impact of trainings on employees' real-life responses to phishing emails, three simulated phishing emails were sent to all study participants. We collaborated with two security experts who were in charge of phishing campaigns at the organization to design these tests with the themes of ``Email client upgrade'', ``Data breach'', and ``Security alerts''. Between 20 and 50 days after answering Q1, all participants received three phishing tests from the IT team. There were intervals of 6-7 days between each test, and participants received each test simultaneously. If a participant clicked the link within the phishing email, they would be directed to a webpage indicating that ``you clicked on a simulated phishing test''. We included one example of a phishing email and the webpage in the Supplementary Material. The tests were distributed in a sequence of easy, moderate, and difficult to identify as phishing\footnote{Two researchers ranked the emails with the phishing scale \cite{steves2020categorizing}, while one IT security expert relied on their experience with simulated phishing tests. They independently assessed and reached mutual agreement on the difficulty of identifying the three phishing emails.}. Refer to figure \ref{fig-1} for an illustration of the study procedure.

\subsection{Measures}

We collected demographic information and participants' current responding ``strategies or practices'' when receiving suspicious emails in Q1. Demographic questions included participants' job positions, the organizational entities they worked for, starting date at the current organization, gender, and age group. Additionally, we employed scales and simulated phishing test records to evaluate the training effects and gathered feedback from participants. 

\subsubsection{Assessment scales}

We included a self-efficacy (\textbf{SE}) scale and support-seeking (\textbf{SS}) scale (adapted version) in our pre- and post-training assessments (Q1, Q2, and Q3). Given the absence of a validated anti-phishing self-efficacy scale, we utilized two dimensions of self-efficacy sourced from studies with good construct validity and reliability. These dimensions evaluate distinct aspects of self-efficacy. Self-efficacy 1 (SE1) measured participants' confidence in learning and updating their knowledge of phishing attack techniques with three items \cite{williams2020developing}, while Self-efficacy 2 (SE2) evaluated participants' confidence in recognizing suspicious emails with four items \cite{ng2009studying}. Meanwhile, we adapted the \textit{Instrumental Support Seeking scale}\footnote{Instrumental Support Seeking scale: assesses the inclination to seek advice, information, and feedback from one's social network during stressful situations \cite{greenglass1999proactive}.} to evaluate participants' intention to seek support when receiving suspicious emails \cite{greenglass1999proactive}. Three researchers reviewed and adapted the SS collaboratively. Afterward, one external expert assessed and confirmed that the adapted SS was more accurate in measuring support-seeking in the phishing context compared with the original scale. We used SE and SS with pilot study participants and received positive feedback. We include SE and SS in the Appendix \ref{appenx3}. 

\subsubsection{Performance metrics to evaluate participants' phishing resilience}

We used the number of reported and non-clicking on links within simulated phishing tests as indicators of participants' resilience to phishing emails. There are mixed findings regarding the effectiveness of phishing campaigns as a form of intervention \cite{lain2022phishing,yeoh2022simulated,hielscher2023employees}. Nonetheless, in agreement with \cite{chaudhary2022developing,reinheimer2020investigation}, simulated phishing tests may be an ecologically valid evaluation method, reflecting participants' natural responses to phishing attempts, if the measures of success are carefully designed, especially in combination with other evaluation methods. We also recorded when participants reported the phishing tests to investigate how quickly they reported a simulated phishing email. We provided all study participants with the same instructions, ``the IT department will send you three simulated phishing tests in the coming month. If you spot any suspicious emails, please report them: forward them as an attachment to report-a-phish@anonymized'' (the standard reporting procedure at the organization).  

\subsubsection{Training feedback} \label{feedbackQ}

In Q2, we asked the participants the following open-ended questions:
\begin{itemize}
    \item What strategies or practices would you apply when receiving suspicious emails?
    \item Which aspects of the training, if any, do you consider useful for learning?
    \item How do you anticipate that this knowledge will help you in your work?
\end{itemize}

In Q3, for the control group, we collected their feedback on the phishing campaigns at the organization with the above three questions. For two treatment conditions, we prepared the following two questions:

\begin{itemize}

    \item Please rate the effectiveness of the training in helping you defend against future phishing attempts. (Select from: Poor, Fair, Good, Very good, Excellent) Optional question: ``\textit{Please enter your comment}''. 
    \item How likely are you to recommend this training to a colleague? (Select from: Very unlikely, Unlikely, Neutral, Likely, Very likely)
\end{itemize}

\subsection{Ethical considerations}
The study design received approval (ERP 22-061) from the ethical review board before data collection. The first author signed a non-disclosure agreement with the organization to access the log data of simulated phishing tests and reporting records for the purpose of this study. We ensured that our study posed no potential harm to the participants. There was no dangerous link to follow, no malicious attachment to be downloaded, and nothing that could lead to a leak of personal information in our simulated phishing tests. The experimental design avoids any emotional distress due to doubts about having really fallen victim to a phishing attack. During the recruitment process, we informed the participants that there were two conditions (in-person and remote) and that they would be randomly assigned to one condition. We were transparent regarding the tasks they would perform in the registration questionnaire. To compensate for participants' time commitment, we offered €25 gift vouchers to each participant of treatment conditions and a €10 gift voucher to each participant of the control group the day after we sent Q3 (even if they did not complete all the questionnaires). We informed participants about their right to opt out of in-situ phishing tests through the information sheet provided for all three conditions and reiterated this at the end of each in-person training session. We only used pseudonymous data in our analysis to protect participants' privacy \cite{resnik2018ethics}. We preregistered the study before launching it\footnote{Pre-registration: \url{https://aspredicted.org/qi8ka.pdf}}.

\subsection{Data collection and analysis}

\subsubsection{Data collection} We conducted six sessions each for group discussion and role-playing training over 19 days in July 2023. We collected 105 complete answers from Q1 (all participants), 70 complete answers from Q2 (all training attendees), and 103 complete answers from Q3 (34 from the control group, 35 from group discussion, and 34 from role-playing). We have non-clicking and reporting records from the three in-situ phishing tests from 105 participants\footnote{In addition to the aforementioned data, we captured video recordings of role-playing sessions and audio recordings of group discussion sessions. For role-playing, we collected 12 phishing emails designed by the participants. Furthermore, we gathered the completed templates for analyzing real phishing emails by group discussion participants. Due to length constraints, the analysis of these collected materials is deferred to future studies.}.

\subsubsection{Preliminary analysis} Prior to our main analysis, we examined the scale validity and randomization of our group assignment. We converted the five-point scale SE1 into a seven-point scale to integrate the two dimensions of SE \cite{statistics2020transforming}. We examined the scales' factor structure and measurement validity following the recommendations of Kline and Schmitt et al. \cite{schmitt2018selecting,kline2011principles}. First, we screened the data for missing entries or errors and evaluated item distribution. The data exhibited a non-normal distribution: 11 items showed skewness beyond the -1 to 1 range, and four items displayed kurtosis exceeding 5. Second, we specified measurement models for both scales and employed the maximum likelihood mean adjusted (MLM) method to estimate model parameters, addressing non-normality. Third, model fit was assessed using the Robust Comparative Fit Index (CFI), Robust Tucker-Lewis Index (TLI), Robust Root Mean Square Error of Approximation (RMSEA), and Standardized Root Mean Square Residual (SRMR). A favorable fit is indicated by CFI and TLI values exceeding .9, and RMSEA and SRMR values below .08 \cite{byrne2013structural,joreskog1993lisrel,kline2011principles}. Fourth, we examined modification indices for models with suboptimal fit to identify areas for enhancement. Only intra-latent covariances were introduced. Fifth, Cronbach's alpha was calculated to assess internal consistency reliability, with values exceeding .70 generally deemed acceptable. Lastly, to examine the randomization of our group assignment, we used the chi-square analysis and the Kruskal-Wallis test to check the distribution of demographic factors among the three groups in Q1. 

\subsubsection{Main quantitative analysis} Given our relatively small dataset, which was not normally distributed and contained several outliers in each condition (refer to the box plot in Appendix \ref{app-box}), we applied non-parametric analysis to the primary study variables (\textbf{SE} and \textbf{SS} scores). Therefore, we used Friedman's one-way repeated measures analysis (non-parametric analogy to repeated-measures ANOVA) for within-groups analysis to investigate whether the trainings have effects on participants \cite{field2013discovering}. We applied the Kruskal–Wallis one-way analysis of variance to compare the training effects between groups \cite{corder2011nonparametric}. We applied the Bonferroni correction to all p-values obtained from post-hoc pairwise comparisons in our tests and present Bonferroni adjusted p-values in the findings \cite{jafari2019and}.

We performed chi-square and Kruskal-Wallis tests to analyze the \textbf{phishing test} results. We conducted regression analyses, with the dependent variables being the sum of non-clicking and the sum of reported to examine whether the performance of non-clicking and reporting was influenced by the measured variables. We applied the non-parametric Mann–Whitney U test to analyze the differences between groups in perceived effectiveness and likelihood to recommend. Main and supporting quantitative data were analyzed and visualized with SPSS 28 and R (lavaan and ggplot2 packages). We provide an overview of main quantitative analysis methods and the corresponding research questions in Table \ref{RQtable}. We include the anonymized dataset and SPSS syntax used for analysis in the Supplementary Material to allow verification and reproducibility.

\begin{table*}[h]
\centering
\caption{Overview of research questions and corresponding quantitative analysis methods.}
\label{RQtable}
\begin{tabular}{@{}p{0.3\linewidth}p{0.35\linewidth}p{0.3\linewidth}@{}}
\hline
\textbf{Research Question} & \textbf{Analysis Method} & \textbf{To Examine} \\
\hline
\hline
\textbf{RQ1}: What is the effect of role-playing training on employees’ anti- &
 Related-samples Friedman's two-way analysis of variance by ranks & whether the trainings have effects on participants (\ref{self-efficacy})\\
\cline{2-3}
phishing \textbf{self-efficacy} compared to group discussion and the control group? & Kruskal–Wallis one-way analysis of variance with Bonferroni-adjusted pairwise z-test for post-hoc analysis & the training effects between groups (\ref{kruskal})\\ 
\hline
\hline
\textbf{RQ2}: What is the effect of role-playing training on employees’ \textbf{support-seeking }intention when  &
 Related-samples Friedman's two-way analysis of variance by ranks &  whether the trainings have effects on participants (\ref{self-efficacy})\\
\cline{2-3}
receiving phishing emails compared to group discussion and the control group? & Kruskal–Wallis one-way analysis of variance with Bonferroni-adjusted pairwise z-test for post-hoc analysis & the training effects between groups (\ref{kruskal}) \\ 
\hline
\hline
\textbf  &
Chi-square Analysis &  whether there are difference in non-clicking and reporting between groups (\ref{reporting}, Appendix \ref{table-6})\\
\cline{2-3}
{\textbf{RQ3}}: How do group discussion and role-playing training influence employees' \textbf{response} to phishing & Kruskal–Wallis one-way analysis of variance with Bonferroni-adjusted pairwise z-Test for post-hoc analysis & the differences in non-clicking and reporting in three simulated tests combined (\ref{reporting}) \\
\cline{2-3}
attacks? & Linear Regression analysis  & the effect of SE, SS, and demographics on the performance of non-clicking and reporting (\ref{linear}) \\ 
\hline
\hline
\textbf{Feedback analysis} &
 Mann–Whitney U test  & whether there are difference in ``perceived effectiveness'' \& ``likelihood to recommend'' (\ref{feedback})\\ 
\hline
\end{tabular}
\end{table*}

\subsubsection{Qualitative analysis}

We conduct a qualitative analysis on the collected feedback (\ref{feedbackQ}) with MAXQDA \cite{maxqda}. The first author read through the feedback, took notes, generated codes from meaningful segments, and organized these codes into preliminary categories \cite{kuckartz2019analyzing}. Subsequently, two other authors reviewed the formulated code system and held two discussion meetings to improve the clarity and precision of the code system. Following this refinement, the first author coded the feedback in MAXQDA. As part of quality assurance, another author examined the coded segments to ensure accuracy and consistency. Afterward, to compare differences in counter-phishing measures among three conditions, we employed MAXQDA to retrieve and visualize the frequency of ``counter practices'' coded in the participants' responses. Regarding the usefulness of the training, two authors thoroughly reviewed the coded segments and categorized them into four distinct themes \cite{clarke2017thematic}. We include the code system and example quotes in the Appendix \ref{appenx4}.  

\section{Quantitative results}
\label{sec-Quanti}

\subsection{Preliminary analysis}\label{preliminary}

\subsubsection{Validity and reliability of the scales} 

The self-efficacy (\textbf{SE}) scale showed a good fit after adding covariance between two items (\(\chi\)²(12) = 17.757, p = .123, CFI = 1.000, TLI = 0.999, RMSEA = 0.0144, SRMR = 0.023): ``I am confident I can recognize a suspicious email'' and ``I am confident I can recognize suspicious email headers''. The reliability of the SE scale was excellent in Q1, Q2, and Q3 (see Table \ref{table-2}). The support-seeking ((\textbf{SS}) ) scale showed an acceptable model fit after adding covariance between two items (\(\chi\)²(19)=38.582, p = .005, CFI = 0.939, TLI = 0.911, RMSEA = 0.091, SRMR = 0.066): ``I try to talk and explain the suspicious elements of an email in order to get feedback from my colleagues'' and ``Before clicking anything within a suspicious email I'll talk with a colleague about it'' (refer to Appendix \ref{app-factor} to see factor loading of both scales). The reliability of the SS scale was good in Q1, Q2, and Q3 (see Table \ref{table-2}). 

\begin{table*}[!ht]
\centering
\caption{Descriptive and Cronbach's alphas (\(\alpha\)) for SE and SS scales.}
\label{table-2}
\begin{tabular}{@{}llllllll@{}}
\toprule
       & n missing & Mean  & SD   & min & median & max & \(\alpha\) \\ 
\midrule
Q1 SE & 0          & 36.32 & 7.45 & 16  & 37     & 49  & 0.91  \\
Q2 SE & 35         & 39.89 & 7.32 & 10  & 41     & 49  & 0.93  \\
Q3 SE & 2          & 39.70 & 7.19 & 11  & 41     & 49  & 0.91  \\
Q1 SS & 0          & 22.23 & 5.05 & 8   & 23     & 32  & 0.83  \\
Q2 SS & 35         & 25.61 & 4.36 & 14  & 26     & 32  & 0.81  \\
Q3 SS & 2          & 24.06 & 5.45 & 12  & 24     & 32  & 0.89  \\
\bottomrule
\end{tabular}
\begin{tabular}{@{}p{\textwidth}@{}}
\smallskip
\centering
\textit{Note:} \textbf{Q1 SE} is the sum of the seven items in Q1, ranges from 7 to 49. Higher scores indicate higher self-efficacy.\\ \textbf{Q1 SS} is the sum of the eight items in Q1, ranges from 8 to 32. Higher scores indicate higher support-seeking intention.
\end{tabular}
\end{table*}

\subsubsection{Randomization check}

The results from the chi-square (\(\chi\)²) analysis found no significant differences between the three groups in terms of gender proportion (\(\chi\)²(6, 105) = 2.850, p = .827), faculty (\(\chi\)²(4, 105) = 6.648, p = .156), and age group (\(\chi\)²(8, 105) = 11.855, p = .158) in Q1. The Kruskal-Wallis test also revealed no significant differences in organizational tenure (H(2) = 2.05, p = .359). 

\subsection{Training effects on SE and SS}\label{trainingeffects}

\subsubsection{\textbf{Training effects compared within groups}}
\label{self-efficacy}

We employed the Related-samples Friedman's two-way analysis of variance by ranks to assess if there are training effects on SE and SS in both trainings (refer to Table \ref{tab-3} to see the full results of the analysis). 

\begin{table*}[!ht]
\centering
\caption{Related-samples Friedman’s two-way analysis of variance by ranks.}
\label{tab-3}
\begin{tabular}{@{}llllll@{}}
\multicolumn{6}{c}{\textbf{Group Discussion} (N = 35)} \\
\midrule
\multicolumn{3}{c}{Self-Efficacy} &
  \multicolumn{3}{c}{Support-seeking} \\
 &
  \(\chi^{2}(2) = 16.924\) &
  Sig. \textless{} .001 &
   &
  \(\chi^{2}(2) = 16.217\) &
  Sig. \textless{} .001 \\
\midrule
\multicolumn{6}{c}{Pairwise comparison} \\
 & Z-stat. & Adj.sig (Sig)* & & Z-stat. & Adj.sig (Sig) \\
\midrule
Q1-Q2 & -2.749 & .018 (.006) &  & -3.287 & .003 (.001) \\
Q1-Q3 & -3.884 & \textless{} .001 &  & -3.167 & .005 (.002) \\
Q2-Q3 & -1.135 & .769 (.256) &  & .120 & 1 (.905) \\
\midrule
\\ 

\multicolumn{6}{c}{\textbf{Role-playing Training} (N = 34)} \\
\midrule
\multicolumn{3}{c}{Self-Efficacy} &
  \multicolumn{3}{c}{Support-seeking} \\
 &
  \(\chi^{2}(2) = 8.835\) &
  Sig. = .012 &
   &
  \(\chi^{2}(2) = 25.878\) &
  Sig. \textless{} .001 \\
\midrule
\multicolumn{6}{c}{Pairwise comparison} \\
 & Z-stat. & Adj.sig (Sig) & & Z-stat. & Adj.sig (Sig) \\
\midrule
Q1-Q2 & -2.304 & .064 (.021) &  & -4.366 & \textless{} .001 \\
Q1-Q3 & -2.425 & .046 (.015) &  & -3.638 & \textless{} .001 \\
Q2-Q3 & -.121 & 1 (.903) &  & .728 & 1 (.467) \\
\bottomrule
\end{tabular}
\begin{tabular}{@{}p{\textwidth}@{}}
\smallskip
\centering
* \textit{Adj.sig (Sig)}: Bonferroni adjusted p-value (unadjusted p-value).
\end{tabular}
\end{table*}

\textit{Immediate training effects (Q1-Q2)}: For group discussion condition, we found statistically significant positive effects on both SE (adjusted p < .001) and SS (adjusted p = .003), comparing the measurements before and after the intervention. For role-playing training, we found statistically significant positive effects on the results of the SS (adjusted p  < .001). However, we did not find significant immediate effects on the SE (adjusted p = .064).

\textit{Day 7 training effects (Q1-Q3)}: For group discussion, we found statistically significant positive effects on both SE (adjusted p = .018) and SS (adjusted p = .005), comparing the measurements before the training and on Day 7. For role-playing training, we also found statistically significant positive effects on the SE (adjusted p = .046) and SS (adjusted p < .001).

\subsubsection{\textbf{Training effects compared between groups}} \label{kruskal}

To compare the Day 7 training effects between the three groups, we analyzed the \textbf{deltas} (score difference between Q3 and Q1) in SS and SE with Kruskal–Wallis one-way analysis of variance. 

\begin{figure}[!ht]
  \includegraphics[width=0.55\textwidth]{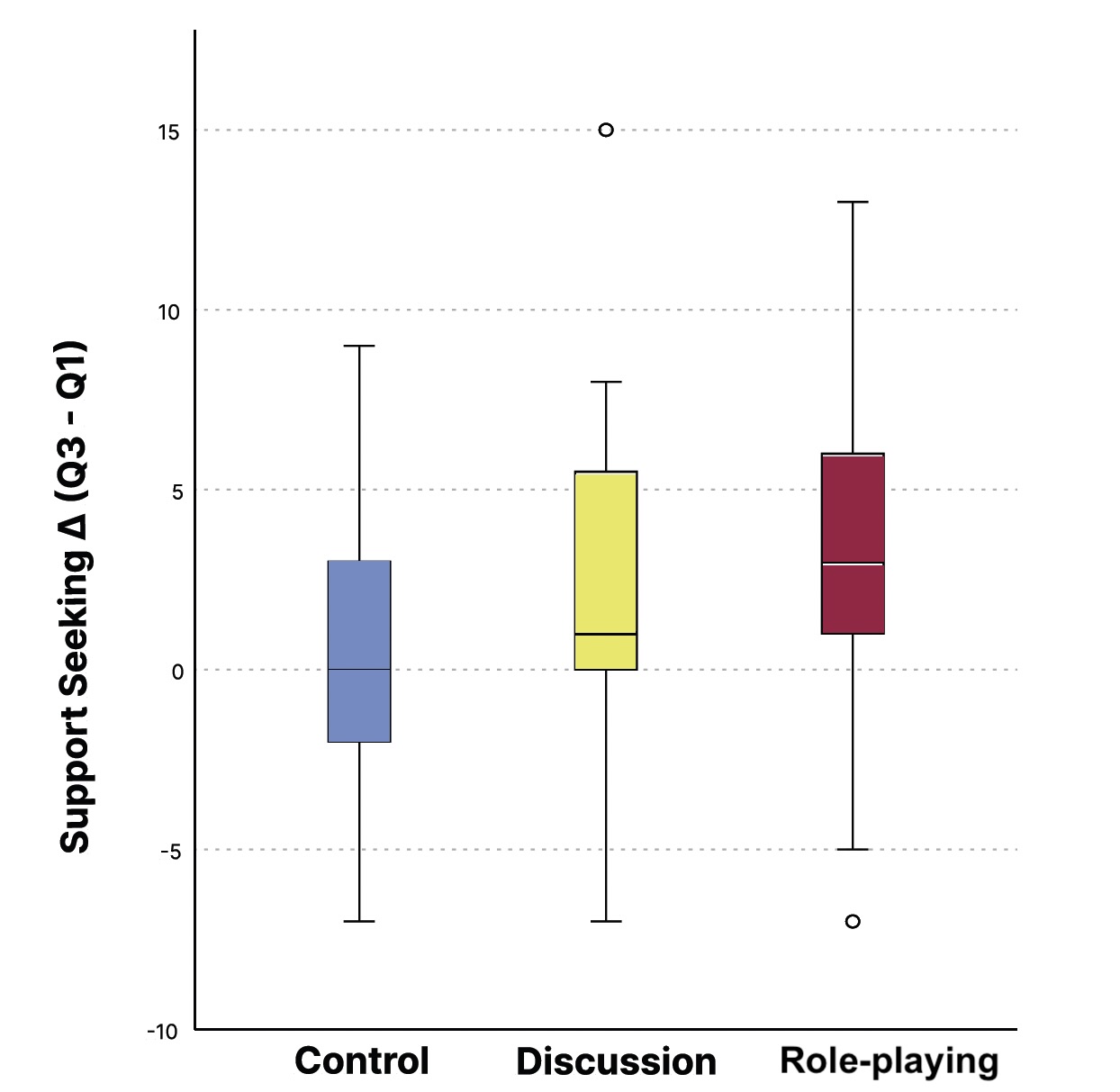}
  \caption{Kruskal–Wallis test of support-seeking deltas.}
  \Description{Figure 2 displays the results of the Kruskal-Wallis test for the deltas of support-seeking intention between Q3 and Q1 when compared across groups. Role-playing training shows the most significant improvement in support-seeking intention compared to the control group. There aren't many differences between two treatment conditions.}
  \label{fig-9}
\end{figure}

\textit{Support-seeking}: The analysis revealed a significant difference between three conditions regarding the deltas of SS (H(2) = 7.169, p = .028), as shown in Figure \ref{fig-9}. The post-hoc analysis (Dunn Pairwise Z-Tests) identified a significant difference in deltas between role-playing and control Group (Z = 2.621, adjusted p = .026). We did not find significant differences between group discussion and role-playing (Z = -.845, adjusted p = 1)\footnote{Similarly, we did not find any statistically significant difference in deltas (Q2-Q1) between two trainings SS (U = 532.5, p = .346), SE (U = 677, p = .467)}.

\begin{figure}[!ht]
  \includegraphics[width=0.55\textwidth]{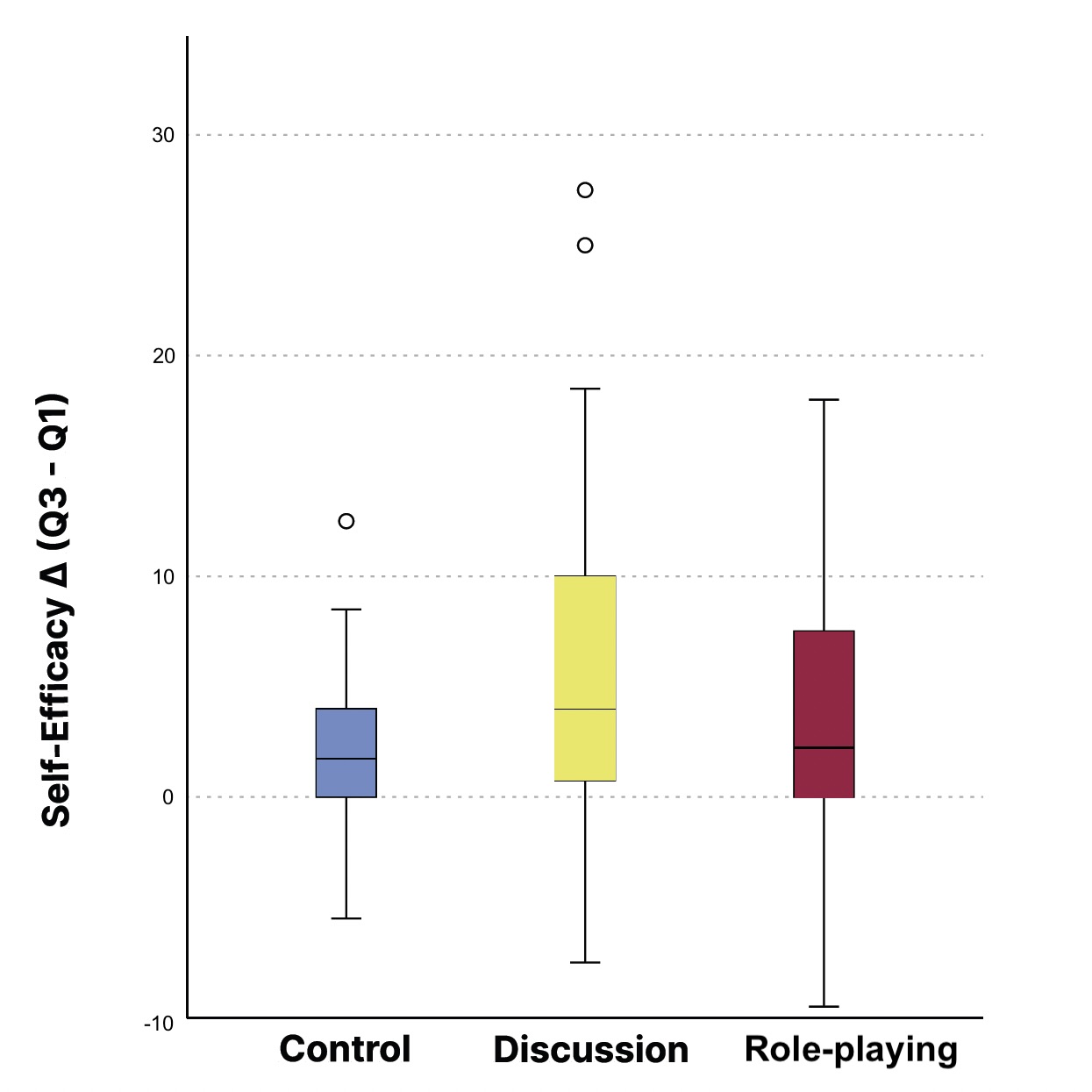}
  \caption{Kruskal–Wallis test of self-efficacy deltas.}
  \Description{Figure 3 shows the results of the Kruskal-Wallis test for the deltas of anti-phishing self-efficacy between Q3 and Q1 when compared across groups. Group discussion shows the most significant improvement in self-efficacy compared to the control group. However, this improvement does not reach statistical significance.}
  \label{fig-8}
\end{figure}

\textit{Self-Efficacy}: The analysis did not reveal a significant difference between any groups regarding the deltas of SE (H(2) = 3.859, p = .145), as shown in Figure \ref{fig-8}. 

\subsection{Phishing test results}\label{performance}
\subsubsection{\textbf{Performance comparison}} \label{reporting}

In Table \ref{table-4}, we present the amount of non-clicking and reported from the three simulated phishing tests. Chi-square analyses for each test revealed no statistically significant difference in non-clicking behavior among the three phishing tests. However, significant differences in reporting behavior were observed across the three tests (see Appendix \ref{table-6}). 

We examined the differences between three conditions in \textbf{non-clicking} (sum) and \textbf{reporting} (sum)\footnote{The non-clicking (sum) represents the total non-clicks on the links of three phishing tests, while reporting (sum) represents the total number of reported phishing tests.} with the Kruskal-Wallis test. The analysis showed no statistically significant differences in non-clicking behavior (H(2) = .002, p = .999), but it did reveal statistically significant differences in reporting behavior (H(2) = 14.662, p < .001).


\begin{table*}[!ht]
\centering
\caption{Number of participants (N) who did not click on the link within the simulated phishing test and reported it to the IT team. Each condition has 35 participants.}
\small 
\begin{tabular}{@{}lcccccc@{}}
\toprule
        & \multicolumn{3}{c}{Non-clicking}                      & \multicolumn{3}{c}{Report-a-phish}                    \\ 
\cmidrule(lr){2-4} \cmidrule(lr){5-7}
        & Client upgrade & Data breach & Security alerts & Client upgrade & Data breach & Security alerts \\
\midrule
Control Group   & 35                    & 33          & 34              & 3                    & 7           & 8  \\
Discussion & 34                    & 34          & 34              & 10                   & 18          & 19 \\
Role-playing & 32                    & 34          & 35              & 4                    & 23          & 20 \\
\bottomrule
\end{tabular}
\label{table-4}
\end{table*}

Post-hoc analysis with Dunn Pairwise Z-Tests revealed statistically significant differences in reporting (sum) between group discussion and control group (Z = 3.235, adjusted p = .004) and between role-playing and control group (Z = 3.391, adjusted p = .002). However, no statistically significant differences were found between two treatment conditions (Z = -.156, adjusted p = 1).

Several participants reported simulated phishing emails within minutes after receiving them. In the first test (Email client upgrade), one participant reported it within one minute, and three participants reported it between 6 and 10 minutes afterward. In the second test (Data breach), five participants reported within one minute, and eight participants reported within 5 minutes. In the third test (Security alerts), five participants reported within one minute, and six participants reported within 5 minutes. 

\subsubsection{\textbf{Linear regression results}}\label{linear}

We estimated linear regression models to examine whether SE and SS predict non-clicking and reporting behaviors. We estimated two separate models for the dependent variables ``non-clicking (sum)'' and ``reporting (sum)'' \footnote{Contrary to our pre-registration, we use non-clicking and reporting as dependent variables, and SE and SS as independent variables. This adjustment is in line with findings from previous studies suggesting that \textit{SE increases intention to report} \cite{kwak2020users}.}, with the independent variables ``SE (Q3)'' and ``SS (Q3)''. As control variables, we included ``working month'', ``gender'', and ``faculties''. The regression results indicated that none of the independent and control variables in both full models had a statistically significant effect on the dependent variables. We provide the regression results in the Appendix \ref{regression}.

\subsection{Feedback analysis}\label{feedback}

The Mann–Whitney U test indicated that there were no significant differences between two trainings in perceived effectiveness (U = 725.5, p = .084), and an analysis of the means revealed high scores for both trainings. Specifically, group discussion had a mean of 4 (CI: 3.74-4.26), and role-playing had a mean of 4.29 (CI: 4.03-4.56). As for the likelihood to recommend (U = 550, p = .527), group discussion had a mean of 4.6 (CI: 4.39-4.81), while role-playing achieved a mean of 4.5 (CI: 4.27-4.73). This suggests that participants found both trainings highly effective and recommendable.

\section{Qualitative findings}
\label{sec-Quali}

\subsection{Changes in counter-phishing practices} 
\label{changes}
In this section, we present a comparative summary of participants' counter-phishing practices as reflected in responses from the three questionnaires (\ref{feedbackQ}). We did not identify noticeable differences among the three conditions when assessing the number of participants who specified specific categories of counter-phishing practices in Q1 (refer to Table \ref{table-3}). Furthermore, we noted only minor variations in the responses provided by the control group in Q1 and Q3. Considering these factors, we adopt the control group as the baseline for comparing coded segments between groups.

\begin{table}[!ht]
\caption{Number of participants (N) who mentioned a specific category of counter-phishing practices; in each cell, a participant is only counted once.}
\begin{tabular}{@{}lcccc@{}}
\toprule
Practices &   & Control & Discussion & Role-playing \\ 
\midrule
Check email header & Q1 & 18 & 20 & 20 \\ 
                   & Q2 & N/A & 18 & 14 \\ 
                   & Q3 & 17 & 18 & 19 \\ 
\midrule
Evaluate email content & Q1 & 14 & 15 & 14 \\ 
                        & Q2 & N/A & 16 & 20 \\ 
                        & Q3 & 10 & 17 & 17 \\ 
\midrule
Do not respond & Q1 & 16 & 12 & 16 \\ 
               & Q2 & N/A & 10 & 8 \\ 
               & Q3 & 19 & 11 & 11 \\ 
\midrule
Block/report & Q1 & 6 & 13 & 11 \\ 
             & Q2 & N/A & 22 & 21 \\ 
             & Q3 & 7 & 18 & 21 \\ 
\midrule
Interact with colleagues & Q1 & 1 & 0 & 1 \\ 
                         & Q2 & N/A & 7 & 7 \\ 
                         & Q3 & 3 & 7 & 9 \\ 
\bottomrule
\end{tabular}
\label{table-3}
\end{table}

\begin{figure}[!ht]
  \includegraphics[width=0.45\textwidth]{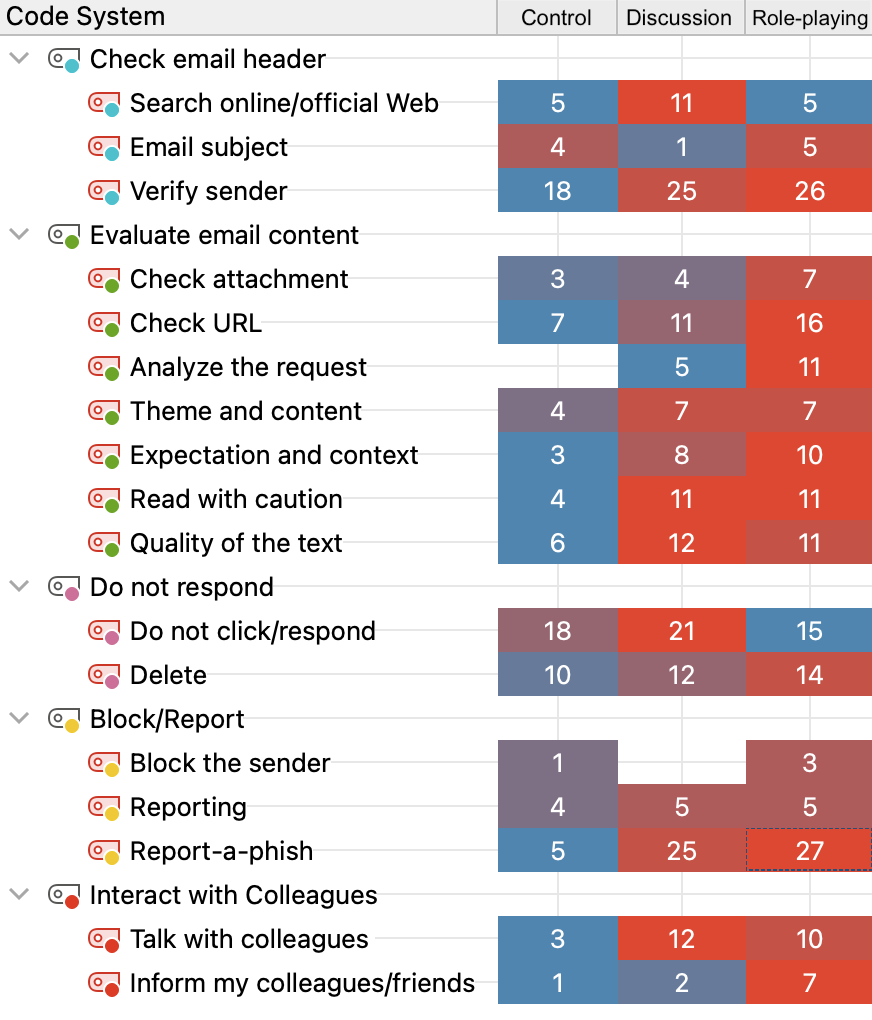}
  \caption{Number of Participants (N) mentioned specific counter-phishing practices across all questionnaires. To enable comparison between the control group and the treatment groups, in each cell, a participant is only counted once, even if they mentioned a topic in multiple questionnaires. Bright red indicates the largest number of the row.}
  \Description{Figure 4 presents the code system and indicates the number of participants who mentioned a specific code in all their questionnaire answers. Compared to Control Group, Discussion Group had more participants mentioning: verify sender, read with caution, quality of the text, do not respond or click, report-a-phish, and talk with colleagues. In comparison to the control group, more participants from role-playing mentioned: verify sender, check URL, analyze request, understand expectation and context, report-a-phish, talk with colleagues, and inform my colleagues/friends.}
  \label{fig-3}
\end{figure}

\subsubsection{Identify phishing with header and content}

More participants of both trainings mentioned checking email details to discern phishing compared to the control group (see Figure \ref{fig-3}). The training attendees mentioned evaluating incoming emails based on ``theme and content'', ``expectation and context'', ``read with caution'', and ``quality of the text'' more than those in the control group. Additionally, more participants of role-playing training indicated reviewing specific email elements, such as ``checking attachments'', ``verifying URL'', and ``analyzing requests'' than group discussion.

\subsubsection{Do not respond}

After the training session, the group discussion generated the highest occurrence of ``do not respond'' subcodes among all groups in Figure \ref{fig-3}. It is noteworthy that a few participants indicated ``do not respond'' and ``report'' at the same time. The response (``do not respond'') does not necessarily indicate participants' exclusive way of responding to phishing emails. 

\subsubsection{Report phishing emails} \label{reportintention}

The number of participants mentioning ``report-a-phish'' (report phishing emails to the IT team) increased noticeably after the training interventions (see Figure \ref{fig-3}). However, only a few participants mentioned their rationale for this change in the questionnaire, including ``informing the IT team'' (P21), ``detecting the phishing attempt'' (P23), as well as ``helping others'' (P53).

\subsubsection{Interact with colleagues}

We noted an increased number of participants indicated talking with/informing their colleagues (friends) regarding phishing emails following the trainings (refer to Figure \ref{fig-3}). This communication often involved \textit{alerting colleagues} about phishing attacks (P11) or seeking support in determining whether the email they received was a phishing attempt (P43).

\subsection{Usefulness of the training}  
\label{usefulness}
In this section, we summarize our findings from analyzing the open-ended question answers regarding ``the usefulness of the training'' and ``how this knowledge will help participants in their work'' (\ref{feedbackQ}). We begin by presenting the utility values that participants attributed to both trainings. Following this, we describe the particular aspects emphasized within each training, the varied levels of enjoyment reported by the participants, and some negative effects of the trainings.

\subsubsection{Utility values of the trainings} Participants gained knowledge of various phishing techniques, communication channels, and attack tactics after both trainings. The training was ``a good reminder of what to look for to assess phishing emails'' (P42). Furthermore, participants emphasized their ``vigilance'' and heightened awareness against phishing threats (P7). Notably, a majority of participants expressed their intention to ``respond with caution'' when receiving emails (P21) and communicate with their colleagues when receiving suspicious emails. The practical examples and group discussion were perceived as useful, for ``they allowed sharing knowledge on the latest phishing attacks'' (P53).

\subsubsection{Learning through group interaction} \label{group-interact} Participants from both trainings found discussion with colleagues useful in deepening their understanding and improving their skills to counter phishing attacks (seventeen participants from group discussion and twelve from role-playing). As P40 of group discussion commented: ``Discussing topics in person helps to remember \textbf{anecdotes and stories from others} that will be helpful in similar future events.'' To clarify, participants in both trainings spent around 40 minutes in discussions. The role-play participants focused on discussing how to craft a phishing email, while the group discussion mainly involved exchanging phishing-related experiences. 


\subsubsection{Thinking like a hacker is useful} \label{hackers}

Fifteen role-playing training attendees referred to the task of designing a phishing email as useful, as it requires participants to \textbf{think from the perspectives of hackers} (P18), identify vulnerabilities of the organization (P11), and examine the elements of phishing emails (P14). Some participants became more aware of the complexity of phishing attacks and attack techniques after role-playing, as P30 indicated:

\textit{Designing phishing email forces us to check how to proceed, and reminds us how the hackers are proceeding/thinking. Really interesting. Adding two normal emails among the phishing emails is really good, as we really have to check and we realized that this is really difficult to see the truth in an email. So this impacted my vision of phishing.}

\subsubsection{Varied levels of enjoyment} \label{interesting}

Seven participants from role-playing training mentioned that the training was \textbf{interesting/fun}, while four participants from the group discussion stressed that it was interesting for them to learn about phishing techniques or analyze phishing emails. In contrast, four participants from both trainings commented that they were not the target recipient of this training. This opinion is associated with their high self-efficacy scores and perceived medium effectiveness of the training (P13 and P68). As exemplified by P19: ``The workshop was excellent in teaching employees about phishing. However, as a computer scientist, I do not know if I was the right target for it.''

\subsubsection{Negative effects of the trainings} \label{ease}

A dozen participants from both trainings lowered their level of self-efficacy after training. Through the questionnaire responses, we get to know that in P8's case, their lowered levels of self-efficacy were due to the realization that it was easier than they thought to create a phishing email and quite difficult to distinguish the phishing email from legitimate ones in the role-playing training, and ``ignoring suspicious emails remains the easiest thing to do, and I cannot ask for support all the time'' (P8). Additionally, in the cases of P27, P36, and P58, they self-reported quite high self-efficacy scores prior to the training and lowered their scores after the training. 

\subsection{Summary of results}

We conducted a mixed-design experiment to assess the effectiveness of role-playing and group discussion training within and between groups. Combining the quantitative and qualitative analysis, we found that: 

\begin{itemize}
    \item RQ1: Group discussion yielded a significant improvement in perceived \textbf{anti-phishing self-efficacy} in both the immediate and Day 7 assessments. The role-playing training did not demonstrate a significant improvement in the immediate assessment but did so for the Day 7 assessment. Both trainings worked similarly well when comparing their effects on anti-phishing self-efficacy on Day 7 (no statistically significant difference in effects between the two trainings). However, these improvements of both trainings did not reach statistical significance when compared to the control group.
    
    \item RQ2: Group discussion and role-playing training significantly enhanced \textbf{support-seeking intention} in the immediate and Day 7 assessments. Again, both interventions worked similarly well when comparing their effects on Day 7 (no statistically significant difference in effects between the two trainings). However, only role-playing training showed statistical significance when compared to the control group. 
 
    \item RQ3: Both trainings were effective in prompting employees to \textbf{report} phishing emails between 20 to 50 days after the training. Employees in the treatment conditions reported phishing tests statistically more often than employees in the control condition. We did not find a difference in \textbf{non-clicking} between groups, but clicking numbers might have been too low to detect differences. In terms of qualitative results, both trainings enhanced employees' vigilance towards phishing emails, increased reporting intention, and promoted their intention to interact with colleagues when receiving phishing emails. Both trainings were perceived as highly effective and were highly recommended by the employees.
\end{itemize}

\section{Discussion}
\label{sec-discussion}

\subsection{Training effects on anti-phishing self-efficacy}

Our research reveals that after participating in a group discussion training, employees perceived higher levels of anti-phishing self-efficacy in the immediate and Day 7 assessment (see \ref{self-efficacy}). When we compare the two trainings, group discussion participants spent more time \textit{sharing knowledge} on the latest phishing attacks and discussing \textit{anecdotes and stories} than in the role-playing training (see \ref{group-interact}). A recent online experiment by Hull et al. might help explain why employees perceived higher self-efficacy after group discussions, finding that anti-phishing training with stories resulted in higher self-efficacy and more accurate phishing detection than training using mindfulness techniques \cite{hull2023tell}. In addition, sharing knowledge is anticipated to reduce the likelihood of information security risks \cite{tamjidyamcholo2014evaluation}. Security stories from others can serve as informal lessons, and these stories impact people's thought processes and corresponding behaviors when making security-relevant decisions \cite{rader2012stories}. Our findings suggest group discussion can be an effective method for delivering rich narrative/story-based anti-phishing training.  

In the immediate assessment, the role-playing training did not yield a significant improvement in anti-phishing self-efficacy. However, participants in the role-playing training showed statistically significant improvements in the Day 7 assessment. This lag points to the importance of measuring effects of anti-phishing interventions beyond the initial effects immediately after an intervention. This lag suggests that participants might not have immediately grasped how role-playing as hackers could be relevant in their professional contexts. Building upon the work of previous role-playing studies \cite{wen2019hack,fatima2019persuasive,becker2023phishing}, our research applies repeated measures to assess the effectiveness of role-playing anti-phishing training, thereby deepening our understanding of its impact.

Self-efficacy is an important step that connects \textit{concordance} and \textit{skills} in the \textit{Security Learning Curve} \cite{sasse2022rebooting}, and is consistently linked to more secure behavior \cite{vance2012motivating}. Previous literature on self-efficacy in cybersecurity often emphasizes measuring employees' self-efficacy, with limited focus on enhancing self-efficacy through cybersecurity training \cite{borgert2023decade,gundersen2022self}. Two previous studies reported that cybersecurity conference attendance and instructor–led training resulted in higher perceived self-efficacy \cite{raineri2020evaluating,khan2023evaluating}. Our findings contribute to the research on phishing interventions by offering empirical evidence that both trainings can be effective methods to boost employees' anti-phishing self-efficacy. Notably, the enhancement of self-efficacy did not reach statistical significance compared to the control group. One plausible explanation could be the presence of a ``ceiling effect'', as a substantial number of employees already possessed high self-efficacy scores prior to the training (see Table \ref{table-2}). 

\subsection{Seeking support when encountering phishing}

Our findings suggest that group discussion and role-playing training demonstrated statistical significance in increasing employees' support-seeking intention when receiving phishing emails in the immediate and Day 7 assessments. While both trainings worked similarly well in elevating employees' support-seeking intention in the Day 7 assessment, only the role-playing training showed significant improvement compared to the control group (see \ref{kruskal}). Role-playing as hackers helped employees scrutinize their organization's vulnerabilities, collaborate with their colleagues, and critically assess the elements of phishing emails (see \ref{hackers}). Some employees may have benefited from this process by grasping the ease with which phishing emails can be created and the difficulty of discriminating these emails from legitimate ones (see \ref{ease}). Such realizations could help dispel the stigmas and shame often associated with being phished \cite{renaud2021shame}, potentially increasing the inclination to seek support when receiving phishing emails. 

Furthermore, the role-playing training we experimented with in this study resembles the ``Red Teaming'' approach, which simulates real-world attacks to fortify organizational resilience against security threats \cite{fenton2016restoring}. Thinking like a hacker enables organizations to anticipate potential threats and take proactive risk reduction measures \cite{esteves2017improve}. Role-playing has been found to be an effective approach to improve phishing detection accuracy and confidence \cite{wen2019hack,sheng2007anti}, enhance phishing awareness, develop phishing knowledge, and stimulate conversations about phishing \cite{baslyman2016smells}. However, only a few previous studies engage participants in playing the role of attackers in social engineering/phishing interventions. Role-playing as attackers facilitates employees' experience of social engineering attacks without deception \cite{beckers2016serious}, educates students about the harms of excessive online information disclosure and the risks of spear-phishing attacks \cite{fatima2019persuasive}, and improves employees' phishing detection abilities \cite{becker2023phishing}. We investigated whether employees can benefit from adopting a hacker's mindset and whether this shift enhances support-seeking intention. As such, we contribute to the literature by demonstrating that role-playing as hackers is an effective approach to enhance employees' support-seeking intention upon receiving suspicious emails.  

We introduced \textit{support-seeking intention} as a useful measurement to assess anti-phishing trainings for several reasons. Given the prevalence of phishing attacks targeting individual email accounts, employees often face these threats in isolation. When employees seek support from their colleagues upon receiving phishing emails, it not only informs the team about the incident but also allows the support-seeker to receive valuable assistance in responding safely to the threat. Furthermore, this support facilitates employees' acquisition of helpful response strategies, developing their ability to handle phishing emails in the future \cite{sasse2022rebooting}. Lastly, cultivating a strong support-seeking intention can pave the way for collaborative efforts in countering phishing, which is a recognized, effective, and indispensable approach for mitigating security breaches within organizations \cite{safa2016human}. 

\subsection{Reporting phishing emails at organizations}

Group discussion and role-playing training increase employees' intention and behavior in reporting phishing emails at organizations (see \ref{reportintention} and \ref{reporting}). Our study supports the notion that reporting can serve as an effective crowd-sourced strategy to counter phishing attacks \cite{lain2022phishing}, as a substantial number of employees reported each phishing test within minutes. Both trainings led to increased numbers of employees mentioning specific counter-phishing practices that they would perform in the future (refer to Figure \ref{fig-3}). This more elaborate process of evaluating phishing emails has been found to link to a greater likelihood of reporting suspicious emails \cite{buckley2023indicators}. 
Employees considered reporting as a means of detecting phishing attempts, informing the IT team of the attacks, and helping others avoid being phished. Various elements of our trainings might be responsible for these favorable outcomes, such as perceiving the severity of phishing attacks by analyzing real phishing emails, which warrant further investigation. In addition to the training approach, sharing reporting statistics might engage employees with simulated phishing campaigns and motivate them to report suspicious emails \cite{hillman2023evaluating}.

Our study contributes to the domain of mitigating phishing attacks at organizations. It is inevitable that employees will interact with phishing emails, considering advanced phishing tactics such as ``spear phishing'' \cite{distler2023influence}. Rather than focusing on the cases where employees interacted with a phishing email, we suggest putting the focus on building a collaborative security culture where employees are encouraged to report incidents proactively. Reporting phishing is one of the most effective methods for the IT team to detect attacks that bypass technical safeguards, and increasing the number of such reports is crucial \cite{althobaiti2021case}. However, previous studies indicate that most employees do not participate in reporting \cite{althobaiti2021case,kwak2020users}. Reporting phishing emails has been used as an indicator to evaluate individuals' responses to phishing and to assess the effectiveness of training \cite{hillman2023evaluating,buckley2023indicators,lain2022phishing}, but it has rarely been considered as one of the objectives of training. In this study, we created two trainings that effectively encourage employees to report phishing emails to the IT team, and were able to empirically demonstrate the effectiveness of these interventions through simulated in-situ phishing tests. 

However, encouraging reporting is insufficient to implement the notion of employees as human firewalls within organizations. Organizations should provide training opportunities to employees who might lack the skills to identify phishing emails. On the other hand, publicly acknowledging reported incidents (e.g., through an employee message board) and validating reported emails are necessary to facilitate reporting as a crowd-sourced defense \cite{jensen2017combating}. Organizations should implement a coherent and consistent protocol for how employees are expected to react to phishing attacks and what response they can expect after reporting. Otherwise, if employees do not perceive the efficacy of their responses, they might feel demotivated to contribute to organizational security \cite{williams2018exploring,herath2009encouraging}.  

\subsection{An enjoyable and effective anti-phishing training}

While our study finds group discussion and role-playing are effective anti-phishing trainings regarding the measurements we evaluated, we identified some constraints when implementing the group discussion approach. As detailed in our training design (see \ref{design}), group discussion was planned for groups of 4-6 participants, allowing all participants to exchange and share their experiences. For groups larger than six participants, it is advisable to split them into two subgroups, which necessitates a spacious room and an expert to facilitate and answer questions in each group. In contrast, for the role-playing training, the ideal subgroup size is 3-4 participants. Thus, even 12 attendees can be divided into three subgroups. As long as the necessary computers, fictitious personas, and email accounts are prepared in advance, a single expert can facilitate and address questions for all groups. While group discussions also serve as an effective and interactive training method, they appear to be more demanding in terms of expert facilitation. 

We are aware of discussions within the security community on the effectiveness and sustainability of anti-phishing trainings, and on the burden they impose on employees, when not generating hidden costs, or even damage, to a company's productivity (\emph{e.g.}, see \cite{lain2022phishing,brunken2023properly}). However, the decision to engage employees in training ultimately resides with the company's management, ideally, in cooperation with the security department. In this case, those in favor of the training have to decide which instrument better serves the purpose of providing some level of preparedness against phishing attacks. In this regard, the choice of what security training to rely upon matters; bad training design choices may offer nothing more than a miserable user experience, providing content that is often \textit{perfunctory} and \textit{arcane}, detached from an employee's practices and expertise \cite{silic2020using}. In short, they are far from being enjoyable and motivating.

The choice of which training to put employees through can therefore make a significant difference in terms of enjoyability, and consequently, engagement, which may result in varied short-term effectiveness. In line with \cite{baslyman2016smells, fatima2019persuasive}, we find that employees consider role-playing \textbf{enjoyable}. More employees in the role-playing group mentioned that the training was interesting than those in the group discussion. When the learning experience is enjoyable, learners are more likely to engage with training and apply what they've learned \cite{hull2023tell,silic2020using}. Thus, we propose using role-playing as an enjoyable and effective approach for anti-phishing training, and potentially for cybersecurity training. (We provide all materials we used in the Supplementary Material to enable others to use and iterate on our approach.)

\subsection{Methodological considerations}

We employed a mixed-design experiment to evaluate the effects of two anti-phishing trainings in the field. Some methodological considerations are relevant for researchers and practitioners conducting social engineering studies. 

\begin{itemize}
    \item \textit{Self-reported change}: Our study assesses the impact of training on participants' perceptions, focusing on two scales, self-efficacy (SE) and support-seeking (SS), changes in self-reported counter-phishing practices, and perceived usefulness of the training. We examined these factors both immediately after the training session and one week later. To the best of our knowledge, we are the first to adapt and measure support-seeking (SS) in phishing intervention studies. 
    \item \textit{Behavioral change}: We also examined behavioral changes after the training through in-situ phishing tests. We utilized non-clicking and reporting behaviors as indicators of employees' phishing resilience. These behaviors directly impact an organization's resistance to phishing attacks.
    \item \textit{Informed consent}: Our findings suggest that it is possible to get meaningful results in social engineering studies with obtaining prior informed consent. Prior informed consent is important for ethical research practices, as it ensures that participants are fully aware of the study objectives and potential risks. However, some researchers and practitioners might worry about prior consent influencing self-reported and behavioral measurements, potentially making participants more alert. In the present study, we used two strategies to counteract the potential influence of informed consent: (a) we standardized the information concerning phishing tests provided for all conditions, and (b) we introduced a waiting period ranging from 13 to 30 days between the compensation email and the first simulated phishing test. In a longitudinal setting, it is likely that participants go back to their typical behaviors even if they know they will receive phishing tests at some point over the course of weeks. The daily work tasks claim participants' attention, enabling the observation of natural behavior without deception. We hypothesize that for social engineering research, which often uses deception \cite{Distler2021systematic}, further longitudinal study designs might lessen the necessity of using deception in certain experimental setups.
\end{itemize}


\section{Limitations and future work}
\label{sec-future}

Our study has a number of limitations. Given that the retention period of training lasts a maximum of five months \cite{jampen2020don}, further examining residual training efficacy in longitudinal field studies is required to evaluate our training. We measure anti-phishing self-efficacy, support-seeking intention, non-clicking and reporting, perceived effectiveness, and likelihood of recommending the training in this study. While self-efficacy has been shown to increase the intention to report phishing emails \cite{kwak2020users}, in our study, neither self-efficacy nor support-seeking intention significantly influenced participants' reporting behaviors when all factors were considered in the regression models. Other factors, such as perceived severity and response efficacy, might merit empirical investigation in future phishing intervention studies.

We recognize that emphasizing self-efficacy in anti-phishing training might lead some employees to become overconfident, potentially reducing their accuracy in detecting phishing emails \cite{das2022evaluating}. Therefore, we incorporated real examples of phishing emails and engaged employees in discussing potential consequences to strengthen their vigilance against phishing threats. In future work, we plan to address the potential side effects of overconfidence bias. Additionally, while stories from colleagues are easily understandable and memorable, the quality and accuracy of such information depend on the narrator's security literacy. Insights from previous studies on ``folk models'' \cite{wash2010folk} and ``security stories sharing'' \cite{wash2011influencing} may illuminate strategies to enhance the effectiveness of group discussion as a training method.

Our study was conducted at a single university, which might have a unique context, organizational culture, and demographic composition compared to other types of organizations. The results could differ when applied to corporate settings, government agencies, or non-profit organizations. Further investigation is needed to assess the applicability of our findings to these diverse contexts. It is possible that we unintentionally attracted tech-savvy participants to register for our study. For future studies, a more systematic sampling method should be devised. Prior informed consent might, to some extent, have influenced our observed results of non-clicking and reporting; we suggest future studies to empirically compare our approaches of counteracting the potential influence brought by informed consent with other creative approaches.

There might be richer insights to be gained from analyzing participants' discussions and phishing email designs, which could shed light on the organization's vulnerabilities and improve anti-phishing training. Due to page constraints, we could not present the findings in this paper. We plan to publish these results in a follow-up paper. 

Organizations deploy phishing campaigns for mainly three objectives: a) examining organizational vulnerability; b) as a form of awareness training; and c) evaluating the effectiveness of an intervention \cite{volkamer2020analysing}. The trade-off between the costs of potential successful attacks and the costs of simulated phishing campaigns is a complex concern. It was outside of this paper's scope to evaluate the costs of training and evaluation measures. In our case, the implementation of in-situ tests entailed considerable coordination and effort from our collaborators and participants. This included four formal meetings and over twenty email correspondences with the Information Security Office to work on the design, testing, and deployment of the simulated phishing emails. This necessitated substantial commitment from the Security Office of both time and expertise, potentially causing interruptions to their workflow. Moreover, we are aware of the time costs and extra workload incurred by our study participants. Future studies should consider these hidden costs associated with deploying phishing tests for evaluation, such as the investment in personnel time and the utilization of IT infrastructure \cite{brunken2023properly}. 

\section{Conclusion}
\label{sec-conclusion}

In this study, we employed a mixed-design experiment to assess the training effects of group discussion and role-playing, involving measurements such as anti-phishing self-efficacy, support-seeking intention, and responses to phishing attacks. Our findings reveal that both trainings were effective in enhancing perceived self-efficacy and support-seeking intention in the Day 7 assessment. However, only role-playing significantly enhanced support-seeking intention compared to the control group. Both trainings contributed to an increase in reporting simulated phishing emails and safe responses to phishing emails. 

Our study contributes to a better understanding of both group discussion and role-playing as effective and interactive anti-phishing training approaches. Also, our study underscores the significance of discussing phishing incidents and sharing anti-phishing practices in the workplace, which can enhance employees' self-efficacy, support-seeking intentions, and vigilance against phishing threats. The study contributes to the field of mitigating phishing attacks by presenting two trainings that effectively prompt employees to report ``phishing emails'' to the IT team. Furthermore, we introduce support-seeking (SS) as a useful measurement to evaluate phishing interventions and devise a promising novel methodology to examine training effects within and between subjects, measuring both self-reported and behavioral changes in the field. Our study demonstrates the feasibility of obtaining informed consent from research participants for simulated phishing tests while still gaining valuable insights from the results.

\begin{acks}
Author 1 acknowledges the financial support of the Institute for Advanced Studies at the University of Luxembourg through a Young Academic Grant (2021). We would like to extend our gratitude to Vincent Koenig and the HCI Research Group, who were instrumental in the conception, design, and pretest of this study. Additionally, we appreciate our colleagues from the Information Security Office, Steve Cannivy and Laurent Weber, for their dedication to the successful realization of this study. We thank the ACs and reviewers for their constructive feedback.
\end{acks}

\bibliographystyle{ACM-Reference-Format}
\bibliography{CHIV1}

\appendix

\section{Self-efficacy and support-seeking scale}
\label{appenx3}

\textbf{Self-efficacy scale}:

Title of scale given to respondents: Self-evaluate confidence

``\textit{Phishing attack is a type of social engineering attack, where attackers send spoofed or deceptive messages to trick a person into revealing sensitive information to the attacker or to deploy malicious software on the recipient’s devices.}

\textit{Please indicate, to what extent, you agree or disagree with the following statements:}''

The following three items are anchored on a 5-point Likert scale, Strongly Disagree (1)/Strongly Agree (5) (from \cite{williams2020developing}):

\begin{enumerate}
    \item It would be easy for me to keep up to date with phishing techniques.
    \item I am able to keep up to date with phishing techniques.
    \item I feel confident in my ability to keep up to date with phishing techniques.
\end{enumerate}

The following four items are anchored on a 7-point Likert scale, Disagree (1)/Agree (7) (from \cite{ng2009studying}):

\begin{enumerate}
    \item I am confident I can recognize a suspicious email.
    \item I am confident I can recognize suspicious email headers.
    \item I am confident I can recognize suspicious email attachment filenames.
    \item I can recognize a suspicious email attachment even if there was no one around to help me.
\end{enumerate}

\textbf{Support-seeking scale} (adapted from \cite{greenglass1999proactive}) :

Title of scale given to respondents: Decision-making in countering phishing

``\textit{Imagine that you have just received a suspicious email in your work account. Please indicate, to what extent, you agree or disagree with the following statements}.''

Respondents are presented with four alternatives: ``not at all true'', ``barely true'', ``somewhat true'', and ``completely true.''

(In scoring responses, 1 is assigned to ``not at all true, 2 to ``barely true'', 3 to ``somewhat true'' and 4 to ``completely true''.)

\begin{enumerate}
    \item When receiving a suspicious email other people's advice can be helpful.
    \item I try to talk and explain the suspicious elements of an email in order to get feedback from my colleagues.
    \item Information I get from others has often helped me deal with suspicious emails.
    \item I can usually identify people who can help me when dealing with suspicious emails.
    \item I ask others what they would do when they receive a suspicious email.
    \item Talking to others can be really useful because it provides another perspective on properly handling suspicious emails.
    \item Before clicking anything within a suspicious email I'll talk with a colleague about it.
    \item When I am in doubt of an email I can usually find a solution with the help of others.
\end{enumerate}

\section{Coding system for qualitative analysis}
\label{appenx4}

\textbf{A. Coding system for changes in counter-phishing practices}

1. \textbf{Check email header}: Participants check email header elements to verify the email.

1.1 Search online/official website: Participants search online for the mentioned organization/stakeholder in suspicious emails to decide whether it's a phishing.

\textit{I google the email address that sent the email.} (P93)

1.2 Email subject: Participants check the subject/title of the email.

\textit{If I'm not familiar with the domain, then I will check the sender name and the email subject to get clarity of the message and purpose.} (P11)

1.3 Verify sender: Participants verify the sender's email address, name, and domain to decide whether it is a phish.

\textit{I verify the sender's mail-address and electronic signature.} (P100) \\

2. \textbf{Evaluate email content}: Participants evaluate the email content to decide the legitimacy of the email.

2.1 Check attachment: Participants check the filename of the attachment.

\textit{I verify the extension of any attachment files.} (P19)

2.2 Check URL: Participants check the link included in the email.

\textit{If I have to click a URL, I check the destination of the link before clicking on it.} (P68)

2.3 Analyze the request: Participants check the requests from the incoming email (e.g., content cues).

\textit{I read through content and understand what action is required on my end - any personal information request will be a red flag.} (P20)

2.4 Theme and content: Participants check the topic/theme/content of the email to decide whether it is legitimate.

\textit{My boss has a personal way of writing, so if it's not his style, I will check the email whether it is from him.} (P23)

2.5 Expectation and context: Does the email fulfill the participants' expectations and fit their routine/context?

\textit{Do I know person? Am I expecting email from that person? Do I expect link or file from that person?} (P90)

2.6 Read with caution: Think before clicking and read content carefully before reacting.

\textit{Being much more careful about the time I take to read emails and ensuring I check everything.} (P55)

2.7 Quality of the text: Grammar, spelling, format, and language of the email.

\textit{I check spelling and formatting, graphic design and what is usually done by specific departments.} (P15) \\

3. \textbf{Do not respond}: Participants choose not to respond to the email's request.

3.1 Do not click/respond: Participants mention they do not click/respond to the suspicious email.

\textit{Don't engage, do not click on any links or images.} (P93)

3.2 Delete: The participants indicate that they delete the email.

\textit{In general, I delete them immediately.} (P85) \\

4. \textbf{Block/Report}: Participants choose to block/report the email.

4.1 Block the sender: Participants mention that they block the sender.

\textit{I block the sender.} (P21)

4.2 Reporting: Participants only indicated reporting, but did not specify reporting to their organization.

\textit{I will report the suspicious emails, then delete them.} (P20)

4.3 Report-a-phish: Participants report suspicious emails to the organization's IT team.

\textit{I send it as an attachment to report-a-phish to have it checked professionally.} (P98) \\

5. \textbf{Interact with Colleagues}: Participants mentioned their colleagues.

5.1 Talk with colleagues: Participants indicate that they talk with their colleagues about the phishing email to make a decision.

\textit{I ask colleagues if they received similar emails. }(P58)

5.2 Inform my colleagues/friends: Participants mention that they will inform their colleagues/friends of the phishing email.

\textit{Not only I will alert my colleagues about any phishing email I may receive, I will also report it to our administrator.} (P11)

\textbf{B. Coding system for usefulness of the training}

1. \textbf{Phishing knowledge}: Participants mention that they learned about different types of phishing emails and attack techniques during the training.

\textit{It was useful to learn about different phishing strategies and experiences from colleagues.} (P24)

2. \textbf{Skills for safe responses}: Participants mention that they learned how to identify phishing emails and how to respond to them.

\textit{To be more vigilant and looking at various details to identify such attacks and differentiate between which email is legit and which is an attack.} (P7)

3. \textbf{Enhanced phishing awareness}: Participants mention that they became more aware of phishing attacks and their severity and prevalence after the training.

\textit{I realized the threat is serious, and my information could get compromised much easier than I thought. I realized, as an employee of an institute, I'm a target of interest. I used not to take these stuff seriously, always thinking not me, I'm not a celebrity, or rich. I've never realized being an employee here could make me attractive to hackers. Now I know!} (P67)

4. \textbf{Emphasized reporting}: Participants mention that they will take reporting phishing emails more seriously in the future.

\textit{Be more cautious and report phishing emails more consistently.} (P37)

5. \textbf{Group interaction}: Participants mention that they learned phishing knowledge and skills by discussing with the group.

\textit{I found the workshop helpful and informative enough. I mainly enjoyed working in analyzing cases and discussing them with the group. I learned a lot from others' experiences and how they deal with phishing.} (P51)

6. \textbf{Think like a hacker}: Participants mention that they found the group work of designing phishing emails and thinking like hackers to be useful.

\textit{The exercise was very useful for understanding how hackers and scammers use relevant and specific information to attack us - very enjoyable to work with colleagues.} (P18)

7. \textbf{Interesting/fun}: Participants mention that the training they attended was interesting/fun/enjoyable.

\textit{I find this workshop very useful and interesting.} (P31)

\section{Chi-square analysis of non-clicking and reporting of each phishing test}
\label{table-6}

\begin{table}[!ht]
\caption{Chi-square analysis (\(\chi\)²(2), N = 105) of each phishing test.}
\begin{tabular}{@{}lcccc@{}}
\toprule
 & \multicolumn{2}{c}{Non-clicking} & \multicolumn{2}{c}{Reporting}     \\ 
\cmidrule(lr){2-3} \cmidrule(lr){4-5}
 & Value & Sig. & Value  & Sig. \\
\midrule
Email client upgrade & 3.639 & .162 & 6.036  & .049 \\
Data breach & .520 & .771 & 15.428 & < .001 \\
Security alerts & 1.019 & .601 & 10.246 & .006 \\
\bottomrule
\end{tabular}
\label{table-5}
\end{table}

A significantly lower report rate for ``Data breach'' in the control group (p < .001) was observed. Using the right-tailed probability of the chi-squared distribution function in the post hoc analysis, we found:
\begin{itemize}
  \item A significantly higher report rate for phishing test ``Email client upgrade'' in group discussion condition (adjusted p = .042).
  \item A significantly higher report rate for phishing test ``Data breach'' in role-playing training (adjusted p = .01). 
  \item A significantly higher report rate for phishing test ``Security alert'' in both group discussion and role-playing training (adjusted p = .003)
\end{itemize}

We noticed that ``Email client upgrade'' had the lowest number of reported incidents compared to the other two tests. To ensure data accuracy, we validated the numbers with the security expert responsible for implementing the phishing tests. They confirmed the accuracy of the reporting numbers for ``Email client upgrade'' and proposed two plausible explanations: firstly, employees may have mistaken the email for spam; secondly, many employees may have been on holiday when they received this email.

\section{Linear Regression Results}\label{regression}

Note that for count data, Poisson regression is typically the preferred method of analysis. As a check, we also conducted Zero-inflated Poisson regression (for reporting) and Quasi-Poisson regression (for non-clicking), both of which led to the same conclusions.

\begin{table}[!h]
\centering
\caption{Linear regression with non-clicking (sum) as the dependent variable.}
\label{regression_nonclicking}
\small 
\begin{tabular}{lcccc}
\hline
\textbf{Variable} & \textbf{Estimate} & \textbf{Std. Error} & \textbf{statistic} & \textbf{p.value} \\ \hline
(Intercept)      & 2.9857  & 0.2399 & 12.447 & <2e-16    \\
Working month    & 0.0011  & 0.0008 & 1.345  & 0.182     \\
Gender male      & 0.0219  & 0.0794 & 0.275  & 0.784     \\
Gender non-binary& 0.1759  & 0.3490 & 0.504  & 0.615     \\
Admin            & 0.1552  & 0.1384 & 1.121  & 0.265     \\
Other faculties  & 0.0335  & 0.0903 & 0.371  & 0.711     \\
Q3SE             & -0.0041 & 0.0053 & -0.775 & 0.440     \\
Q3SS             & -0.0006 & 0.0066 & -0.094 & 0.926     \\ \hline
\end{tabular}
\end{table}

\begin{table}[!h]
\centering
\caption{Linear regression with reporting (sum) as the dependent variable.}
\label{regression_report}
\small
\begin{tabular}{lcccc}
\hline
\textbf{Variable} & \textbf{Estimate} & \textbf{Std. Error} & \textbf{statistic} & \textbf{p.value} \\ \hline
(Intercept) & -0.3275 & 0.7603 & -0.431 & 0.6677 \\
Working month    & 0.0037  & 0.0026 & 1.432  & 0.1557 \\
Gender male    & -0.0287 & 0.2516 & -0.114 & 0.9095 \\
Gender non-binary    & -0.1548 & 1.1063 & -0.140 & 0.8890 \\
Admin      & 0.6438  & 0.4388 & 1.467  & 0.1458 \\
Other faculties  & 0.1059  & 0.2861 & 0.370  & 0.7121 \\
Q3SE       & 0.0295  & 0.0167 & 1.761  & 0.0817 \\
Q3SS       & -0.0013 & 0.0210 & -0.064 & 0.9494 \\ \hline
\end{tabular}
\end{table}

\onecolumn 

\section{Box Plot of Self-Efficacy and Support-seeking}
\label{app-box}

\begin{figure*}[!ht]
  \includegraphics[width=0.8\textwidth]{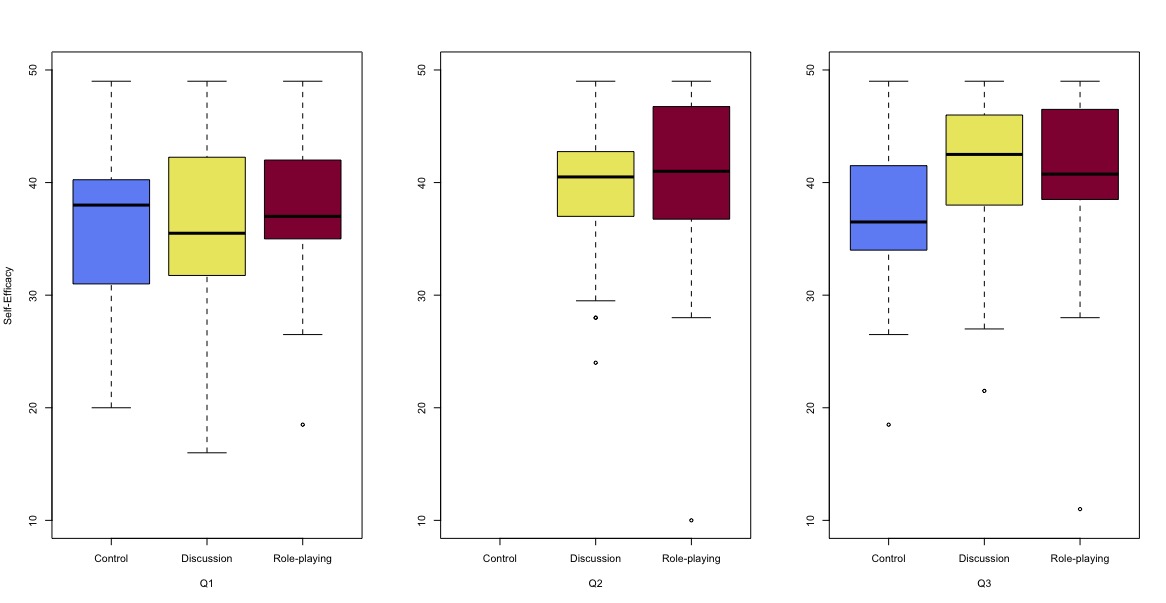}
  \caption{Box plot of self-efficacy scores.}
  \Description{Figure 5 is the box plot of self-efficacy for the three groups in Q1, Q2, and Q3. This figure indicates that there are outliers in each assessment, and the data is not normally distributed.}
  \label{fig-6}
\end{figure*}

\begin{figure*}[!ht]
  \includegraphics[width=0.8\textwidth]{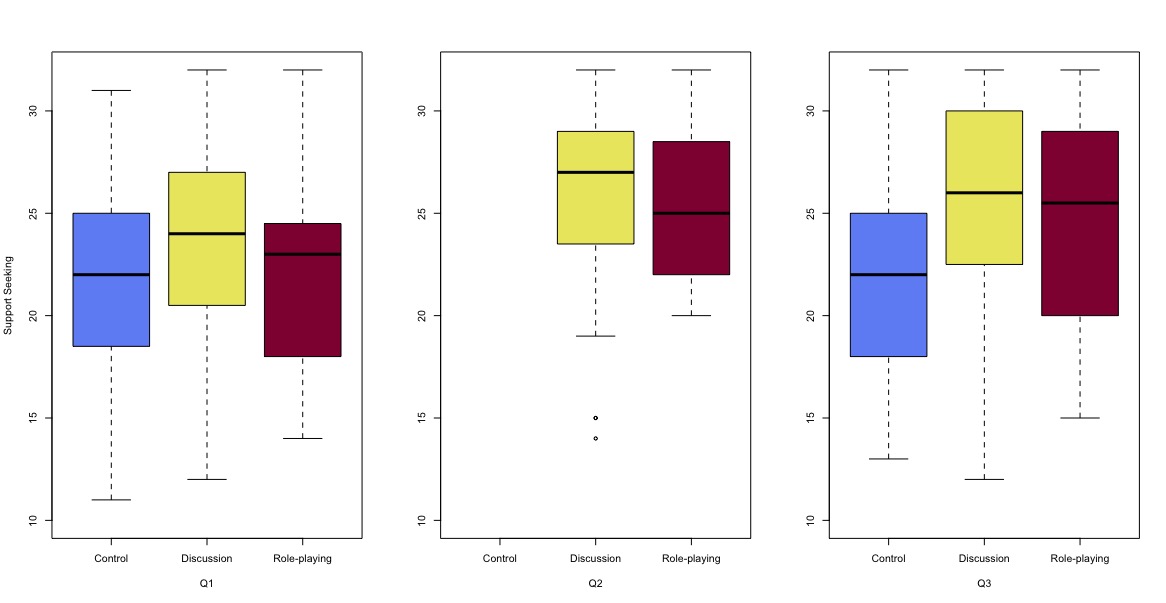}
  \caption{Box plot of support-seeking scores.}
  \Description{Figure 6 is the box plot of support seeking intention for the three groups in Q1, Q2, and Q3. This figure indicates that there are outliers in each assessment, and the data is not normally distributed.}
  \label{fig-7}
\end{figure*}

\section{Factor loading for self-efficacy and support-seeking scales}
\label{app-factor}

\begin{figure*}[h]
  \includegraphics[width=0.9\textwidth]{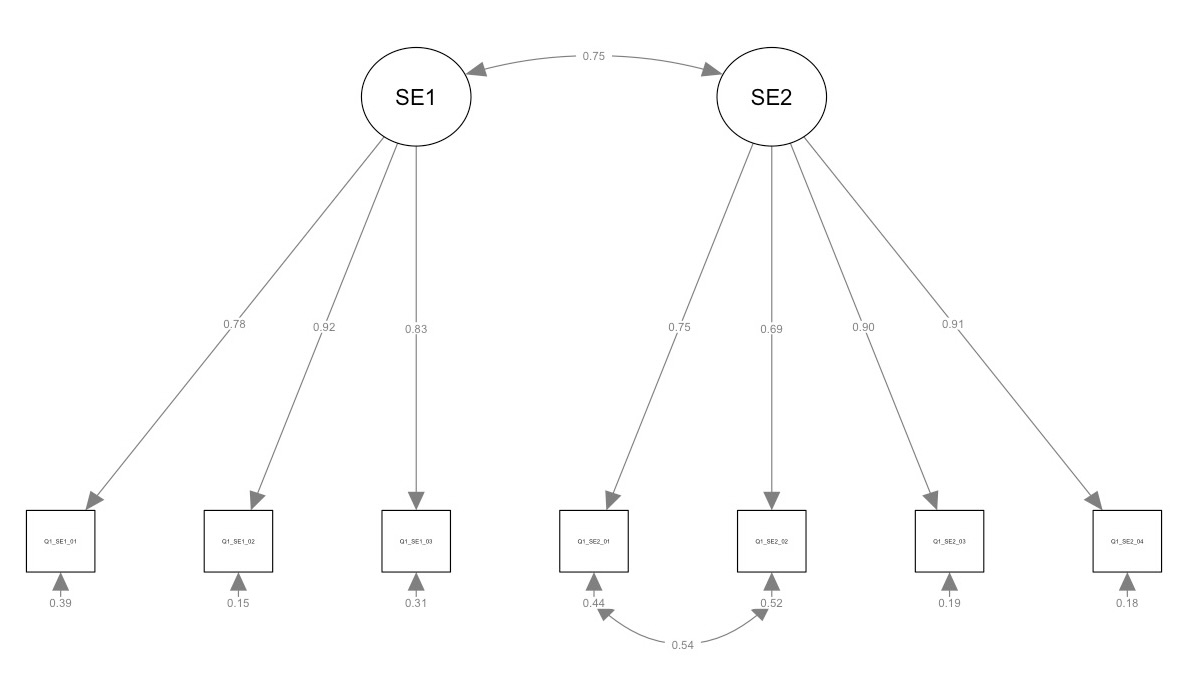}
  \caption{Factor loading for self-efficacy scale items.}
  \Description{Figure 7 is the factor loading for the self-efficacy scale. The model's fit was improved by adding a covariance between two specific items, SE2 item 1 and SE2 item 2.}
  \label{fig-10}
\end{figure*}

\begin{figure*}[h]
  \includegraphics[width=0.9\textwidth]{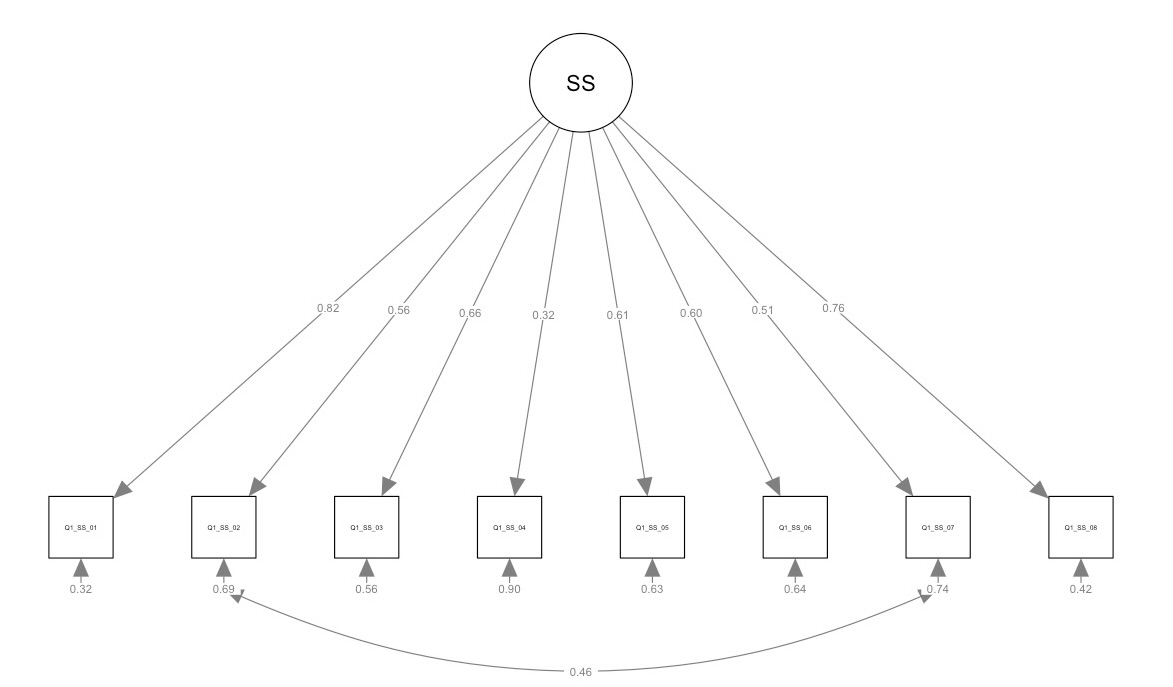}
  \caption{Factor loading for support-seeking scale items.}
  \Description{Figure 8 is the factor loading for the support seeking intention scale. The model's fit was improved by adding a covariance between two specific items, item 2 and item 7. }
  \label{fig-11}
\end{figure*}

\end{document}